\documentclass[pre,amssymb,showpacs]{revtex4}
\usepackage{graphicx}
\usepackage{color}
\usepackage{epsfig}
\usepackage{latexsym}
\usepackage{bm}
\usepackage{ulem}
\usepackage{color}

\begin{document}

\renewcommand{\ni}{{\noindent}}
\newcommand{\dprime}{{\prime\prime}}
\newcommand{\be}{\begin{equation}}
\newcommand{\ee}{\end{equation}}
\newcommand{\bea}{\begin{eqnarray}}
\newcommand{\eea}{\end{eqnarray}}\newcommand{\nn}{\nonumber}
\newcommand{\bk}{{\bf k}}
\newcommand{\bQ}{{\bf Q}}
\newcommand{\q}{{\bf q}}
\newcommand{\s}{{\bf s}}
\newcommand{\bN}{{\bf \nabla}}
\newcommand{\bA}{{\bf A}}
\newcommand{\bE}{{\bf E}}
\newcommand{\bj}{{\bf j}}
\newcommand{\bJ}{{\bf J}}
\newcommand{\bs}{{\bf v}_s}
\newcommand{\bn}{{\bf v}_n}
\newcommand{\bv}{{\bf v}}
\newcommand{\la}{\langle}
\newcommand{\ra}{\rangle}
\newcommand{\dg}{\dagger}
\newcommand{\br}{{\bf{r}}}
\newcommand{\brp}{{\bf{r}^\prime}}
\newcommand{\bq}{{\bf{q}}}
\newcommand{\hx}{\hat{\bf x}}
\newcommand{\hy}{\hat{\bf y}}
\newcommand{\bS}{{\bf S}}
\newcommand{\cU}{{\cal U}}
\newcommand{\cD}{{\cal D}}
\newcommand{\bR}{{\bf R}}
\newcommand{\pll}{\parallel}
\newcommand{\sumr}{\sum_{\vr}}
\newcommand{\cP}{{\cal P}}
\newcommand{\cQ}{{\cal Q}}
\newcommand{\cS}{{\cal S}}
\newcommand{\ua}{\uparrow}
\newcommand{\da}{\downarrow}

\def\lsim {\protect \raisebox{-0.75ex}[-1.5ex]{$\;\stackrel{<}{\sim}\;$}}
\def\gsim {\protect \raisebox{-0.75ex}[-1.5ex]{$\;\stackrel{>}{\sim}\;$}}
\def\lsimeq {\protect \raisebox{-0.75ex}[-1.5ex]{$\;\stackrel{<}{\simeq}\;$}}
\def\gsimeq {\protect \raisebox{-0.75ex}[-1.5ex]{$\;\stackrel{>}{\simeq}\;$}}

%\today

\title{Symmetric exclusion processes on a ring with moving defects}

\author{Rakesh Chatterjee{$^{1, 2}$} } 
\email{rakeshch@fis.unam.mx}
\author{Sakuntala Chatterjee{$^3$}} 
\email{sakuntala.chatterjee@bose.res.in}
\author{Punyabrata Pradhan{$^3$}}
\email{punyabrata.pradhan@bose.res.in}

\affiliation{ $^1$ The Institute of Mathematical Sciences, CIT Campus, Taramani, Chennai 600113, India. \\ $^2$ Instituto de Ciencias F\'isicas, Universidad Nacional Aut\'onoma de M\'exico, Cuernavaca 62210, M\'exico  \\ $^3$Department of Theoretical Sciences, S. N. Bose National Centre for Basic
Sciences, Block-JD, Sector-III, Salt Lake, Kolkata 700106, India. }

\begin{abstract}

\noindent{We study symmetric simple exclusion processes (SSEP) on a ring in 
the presence of uniformly moving multiple defects or disorders - a 
generalization of the model proposed earlier [Phys. Rev. E {\bf 89}, 
022138 (2014)]. The defects move with uniform velocity and change the particle 
hopping rates locally. We explore the collective effects of the defects 
on the spatial structure and transport properties of the system.
We also introduce an SSEP with ordered sequential (sitewise) update and 
elucidate the close connection with our model.}

\end{abstract}


\maketitle

\section{Introduction}

What happens when a system of interacting particles is subjected to a
time-periodic forcing, which vanishes on average over a full cycle? Do the
particles show directional motion and, if so, in which direction? These
questions have recently attracted a lot of attention in the context of quantum
or classical pumps \cite{Thouless, Switkes, Watson, Astumian, Rahav, Marathe},
Brownian ratchets \cite{Julicher, Marchesoni, Seifert} and single-file motions
of interacting colloids in the presence of ac driving forces 
\cite{Dhar_PRL2007, Dhar_JSTAT2008, Dhar_EPL2011, Dhar_PRE2015}, etc. On a 
more general ground, the effect of time-dependent potential on 
many-particle systems is of great interest in the context of externally 
stirred fluids, such as in microfluidic devices or in an assembly of nano-
particles which could be driven by a periodically moving
external potential \cite{Libchaber, Quake_RMP2005,
Lipowsky_Science1999, Marr_Science2002, Penna_Tarazona_JChemPhys2003,
Tarazona_Marconi_JchemPhys2008}. In fact, recently, colloids in a moving 
optical trap have been of considerable theoretical interest \cite{Zon1, Zon2, 
Pal} and have consequently been investigated in several experiments to 
understand static as well as dynamical aspects of fluctuations in 
nonequilibrium systems \cite{Wang, Blickle, Bechinger, Ciliberto}.

We consider the effect of periodically moving external potential, on a system of 
hard-core particles diffusing on a one dimensional lattice of length $L$, 
in a set-up of symmetric simple exclusion process (SSEP). Without any such 
potential, one has an SSEP with fixed (independent of space and time) 
hopping rates for the particles, which is an old problem \cite{Spitzer} 
and has been intensively studied in the last few decades \cite{reviews, Spohn, Schutz, Santos, Derrida_2001, Karger, Nagar, Rolland, Sadhu, Hegde,
Mallick}. In the presence of a time-periodic potential, the hopping rates 
become explicit functions of time, which has not been explored much until 
recently \cite{Dhar_PRL2007, Dhar_JSTAT2008, Chatterjee2014}.  
Motivated by the recent experiments involving moving optical potential on 
colloidal particles \cite{Bechinger}, we model the external potential as 
`defects' or disorders residing at particular sites where local 
diffusivity is different from the bulk of the system. 
These defects move around the periodic lattice with a fixed velocity 
$v$, such that after a time interval $L/v$ they complete one cycle, thus 
producing a time-periodic potential which drive the system away from 
equilibrium. In our earlier work \cite{Chatterjee2014}, we had considered 
a single such defect which was shown to give rise to a
traveling density wave, having a peak and a trough around the 
instantaneous defect-position and thus generating a particle current in 
the system. As the defect velocity $v$ and the particle density $\rho$ are 
varied, the particle current was shown to have rich behaviors, such as 
current reversal and multiple peaks \cite{Chatterjee2014}.

In this paper, we consider multiple defects, all moving with the same velocity $v$, and investigate how their collective effects influence the transport properties of the system. For simplicity, we mainly focus on two defects 
and our results can be generalized to arbitrary number of defects.
When the defects are placed far apart in the system, 
they act independently and their influence on the system can be
described by using the results for the single-defect case \cite{Chatterjee2014}.
However, when the defects are close to each other, the density patterns created individually around the single defects now start overlapping,
which gives rise to interesting collective effects in the system. 
In particular, when the defects occupy next nearest neighbour sites on the 
lattice, a rather complex spatial pattern, with multiple peaks and troughs, 
emerges in the traveling density wave. Another remarkable effect is 
observed when the defects are closer further, i.e., when they are located 
in the nearest neighbor sites. In that case, the current in the system shows 
particle-hole symmetry for large defect velocity, although this symmetry was 
known to be violated in many of the earlier studies \cite{Dhar_PRL2007, 
Dhar_JSTAT2008, Dhar_PRE2015}, including the single-defect case mentioned 
above \cite{Chatterjee2014}. Despite these qualitative differences, some of 
the interesting broad features 
obtained in our earlier work \cite{Chatterjee2014}, such as polarity reversal 
and multiple peaks in particle current upon variation of $v$ and $\rho$, still 
persist. We also discuss close connection between our model
and an SSEP with sitewise ordered sequential update.

The organization of the paper is as follows. In the next section we define 
the model. In Sec. \ref{sec-3}  we describe our results for multiple defects,
with subsections A and B devoted to the cases when the defects are at 
nearest and next nearest neighbor positions, respectively. Subsection C
contains results for the case when the defects are further apart. 
In Sec. \ref{sec-4} we discuss SSEP with ordered sequential update and present
our conclusions in Sec. \ref{sec-5}.

\section{Model}
\label{sec-2}

\begin{figure}[h]
\begin{center}
\includegraphics[scale=0.6]{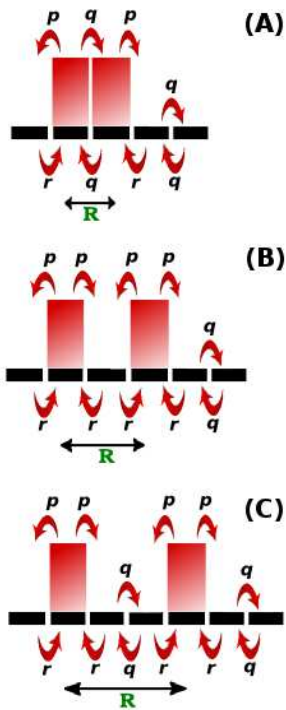}
\caption{(Color online) Schematic representation of the model. Two defects 
reside at two sites which are separated by a distance $R$ as shown in (A) $R=1,$ 
(B) $R=2,$ (C) $R=3.$  Particle hopping is possible from a particular site if the 
site is occupied and the destination site is empty. The hopping rate from a
defect site to a bulk site is $p$ and the reverse process occurs with rate $r$
(for barrier-type defect, $p > r$; for trap-type defects, $p < r$). All other 
hopping rates are $q$. A defect resides at a 
particular site for a residence time $\tau=1/v$ and then moves to the nearest 
neighbour site, say, in the clockwise direction.}
\label{fig-model}
\end{center}
\end{figure}

The model is defined on a one dimensional periodic lattice of $L$ sites where
any site $i= 1, 2, \dots L$ can be occupied by at most one particle. A particle
can hop from an occupied site to one of the neighbouring empty site, with one of the following rates: (i) the hop rate from any of the
defect sites to any the bulk sites is $p/2$, (ii) from one bulk site to a defect 
site is $r/2$ and (iii) all other hop rates
are $q/2$ (see Fig. \ref{fig-model}). Each defect resides at any particular site
for a residence time $\tau=1/v$ where $v$ is the velocity of a defect. Without
any loss of generality, we assume $v>0$, i.e. at the end of the residence time
the defect moves one site to the right.

When the defect velocity vanishes, $v=0$, the hop rates satisfy detailed balance.
Consequently, the ratio between the density $\rho_{i'}$ at a defect site $i'$
and the density $\rho_i$ at a bulk site $i$ is given by $\rho_{i'}/\rho_i =
\exp(-\beta V_0)$, with $\beta V_0 = \ln(p/r)$. Clearly, if $p>r$, the defect
site has a lower density, ($\rho_{i'} < \rho_i$) and, if $p<r$,
density at the defect site is higher than the bulk ($\rho_{i'} > \rho_i$); 
finite $p$ and $r=0$ (finite $r$ and $p=0$) corresponds to an infinite 
potential barrier (well).

For nonzero velocity $v$, the system is driven out of equilibrium and, after a
long time, it reaches a time-periodic steady state where averages are periodic
functions of time, with period $L/v$. In this state, a
nontrivial density pattern, in the form of a traveling wave, emerges.  
For simplicity, in this paper, we mainly analyze the case with only two
identical defects, both moving with the same velocity $v$, separated by 
a distance $R$. Note that $R=0$ corresponds to the case
when both the defects are present on the same site, i.e. only one defect site is
present in the system. This case has already been studied in Ref.
\cite{Chatterjee2014}. We will separately consider three situations, 
(a) $R=1$ when
the defects are present at adjacent sites, (b) $R=2$, the defect sites are next
nearest neighbors and (c) $R \geq 3$, when the defect sites are further apart.
 Our analysis can be easily extended to larger number of defects.

\section{Multiple moving defects}
\label{sec-3}

The configuration of the system is specified by the occupancy of each site and
the position of the defect sites. Let $\{\eta_i^{\alpha_1, \alpha_2}\}$ denote
the occupancy variable  that takes the value $1$ ($0$) if the $i$-th site is
occupied (empty) when the two defects are located at sites $\alpha_1$ and
$\alpha_2$. Starting from
an initial configuration, the system eventually reaches a 
time-periodic steady state where the averages, such as the local density
$\rho_i^{\alpha_1, \alpha_2} (t) = \langle  \eta_i^{\alpha_1, \alpha_2} (t)
\rangle$ at any site $i$, are  periodic functions of time, with a period $L/v$.
The residence time of the defect at a site is $\tau=1/v$ and 
from now onwards, unless stated otherwise, we make all the measurements
only at discrete times $t=m \tau$  ($m=0, 1, \dots \infty$), 
when the defects are about to leave a site and move onto the next one. 
The density pattern created around each defect site also moves with the defect
with the same velocity $v$. Therefore, at any time $t$, the time
evolution of the local density $\rho_i^{\alpha_1, \alpha_2} (t)$ can 
be written in a matrix form as given below,
\be
\langle \rho^{\alpha_1+1, \alpha_2+1} (t+\tau)| = \langle \rho^{\alpha_1,
 \alpha_2} (t)| {\cal W}^{\alpha_1+1, \alpha_2+1},
\label{TW1}
\ee   
where $i$-th element of the row vector, 
\be 
\langle \rho^{\alpha_1, \alpha_2} (t)| \equiv \{ \rho_1^{\alpha_1,\alpha_2}(t), 
\dots, \rho_i^{\alpha_1,\alpha_2}(t), \dots \rho_L^{\alpha_1,\alpha_2}(t) \},
\ee 
denotes the local density at site $i$ and the elements of the time-evolution
operator ${\cal W}^{\alpha_1, \alpha_2}$ is the transition matrix provided that
the defects reside at sites $\alpha_1$ and $\alpha_2$. By definition, in the
time-periodic steady state, density $\rho_{st, i}^{\alpha_1, \alpha_2}$ at site
$i$ has the following property 
\be
\rho_{st, i+1}^{\alpha_1+1, \alpha_2+1} = \rho_{st, i}^{\alpha_1, \alpha_2},
\label{condition-general}
\ee
which will be used later to the find the exact structure of the density profile.
Similarly, the time evolution equations for $n$-point correlations $\langle
\eta_i^{\alpha_1, \alpha_2} (t) \eta_{i+1}^{\alpha_1, \alpha_2} (t) \dots
\eta_{i+n-1}^{\alpha_1, \alpha_2} (t) \rangle$ and their steady-state profile
can be constructed. However, it is not easy to analytically solve
the full Bogoliubov-Born-Green-Kirkwood-Yvon (BBGKY) hierarchy of the 
coupled equations involving $n$-point spatial correlations.
Therefore, in this paper, we confine ourselves to only the mean-field analysis
of Eq. \ref{TW1}, which in fact captures the broad features of the model quite
well.

Our immediate task now is to obtain the elements of the transition matrix ${\cal
W}^{\alpha_1, \alpha_2}$. As we shall see in the following sections, the
transition matrix elements actually depend on two-point correlations and, within
mean-field approximation, determines the structure of the traveling density
wave and the particle current in a self-consistent way. 
For the most part of the paper, we focus on the case of 
infinitely large potential barrier ($r=0$).   

%This limit is amenable to the 
%mean-field theoretical studies. We also show that in the more general case when  
%which capture quite well the broad features also
%in the general case where all three hopping rates $p$, $q$ and $r$ are nonzero.
%Moreover, we later discuss the case with $r \ne 0$ and $p=q=0$ (defects acting
%as potential well) which is actually related to the case of defects with a
%potential barrier by a particle-hole transformation (see section III.B). 

\subsection{$R=1$: two nearest neighbor defect sites with infinite potential
 barrier}

Let us denote the defect positions at one particular configuration as
$\alpha_1=\alpha$ and $\alpha_2=\alpha+1$. We first consider the case when there is
no bulk hopping in the system, i.e. $q=0$. Since infinite potential barrier
corresponds to $r=0, p \ne 0$, a particle can only hop out of a
defect site into the bulk. Two defect sites being in the adjacent positions,
this means a particle can hop only from site $\alpha$ to site $\alpha-1$
and from site $\alpha +1$ to site $\alpha+2$. In other words, only the
following two transitions (each occurring with rate $p$) 
are possible in the system: (a)  $0\hat{1} \hat{X} \rightarrow 1 \hat{0}
\hat{X}$ and (b) $\hat{X} \hat{1} 0 \rightarrow \hat{X} \hat{0} 1$. Here we have
denoted the defect sites by  cap and $X$ can be either $0$ or $1$. 
The transition matrix element ${\cal W}^{\alpha+1, \alpha+2}_{i,j}$, 
the probability that a particle at site $i$
hops to site $j$ within the defect residence time $\tau$, can be
written as,
\bea
\nn 
{\cal W}_{i,j}^{\alpha+1, \alpha+2} &=& 1-a_- \;\;\;\;\;\; i=j=\alpha+1 \nonumber \\
{\cal W}_{i,j}^{\alpha+1, \alpha+2} &=&1-a_+ \;\;\;\;\;\; i=j=\alpha+2  \nonumber \\
{\cal W}_{i,j}^{\alpha+1, \alpha+2} &=& a_+ \;\;\;\;\;\; i=j-1=\alpha+2 \nonumber \\
{\cal W}_{i,j}^{\alpha+1, \alpha+2} &=& a_- \;\;\;\;\;\; i=j+1= \alpha+1 \nonumber \\
\label{eqn-R1}
\eea
where all other off-diagonal elements are $0$ and diagonal elements
are $1$.

Here, $a_+$ ($a_-$) is the conditional probability that a defect site,
provided it is occupied, exchanges particle with its right (left) neighbour
during the time-interval $\tau$. For $R=1$, the expressions for $a_\pm$ is
particularly simple and is given by
\bea
a_{+} &=& \left[ \frac{\langle \eta^{\alpha, \alpha+1}_{\alpha+1}
(1-\eta^{\alpha, \alpha+1}_{\alpha + 2}) \rangle}{\kappa
\langle \eta^{\alpha, \alpha+1}_{\alpha+1}\rangle} \right], \nonumber \\
a_{-} &=& \left[ \frac{\langle (1-\eta^{\alpha,\alpha+1}_{\alpha-1})
\eta^{\alpha,\alpha+1}_{\alpha} \rangle}{\kappa
\langle \eta^{\alpha, \alpha+1}_{\alpha }\rangle} \right],
\label{a+-R1}
\eea
%\normalsize
where $1/\kappa(v) = (1-e^{-p/2v})$ is the rate with which a local 
configuration $0\hat{1} \hat{X}$ or $\hat{X} \hat{1} 0$ goes to
$1 \hat{0} \hat{X}$ or $\hat{X} \hat{0} 1$, respectively, during time $\tau$, 
assuming that the decay process is Poissonian.

\begin{figure}[h]
\begin{center}
\leavevmode
\includegraphics[width=9.2cm]{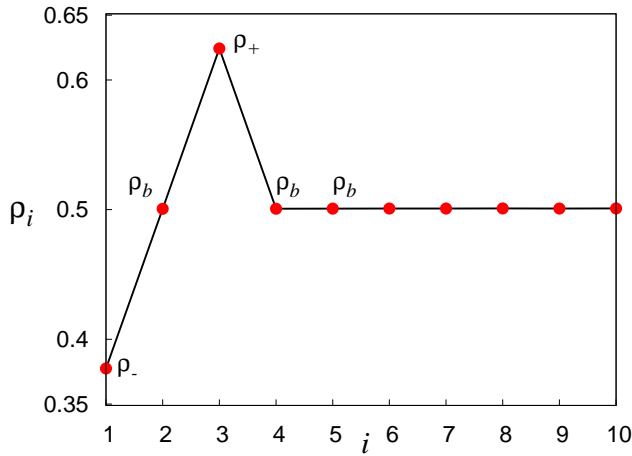}
\caption{(Color online) Time-averaged density profile
of the traveling wave where the two defects are located at two adjacent sites,
with defect velocity $v=1.0$ for system size $L=512$ and hopping rates $p=1$ and 
$q=r=0$. In this figure, the
defects reside at sites $1$ and $2$, the bump is at site $3$ and the trough is
at site $1$. Densities at all other sites are $\rho_b \approx \rho$, which
equals to the global density in the limit of large system size $L \gg 1$. }
\label{x-rho-R1}
\end{center}
\end{figure}

To study the density profile, we consider a particular case when defects 
were at sites $1$ and $2$, and have just moved to the next sites. 
Then the corresponding transition
matrix can be written in terms of $a_+$ and $a_-$ as
\[
{\cal W}^{2, 3} = \left[ \begin{array}{cccccc}
        1 & 0 & 0 & 0 & \dots & 0 \\
        a_- & (1-a_-) & 0 & 0 & \dots & 0 \\
        0 & 0 & (1-a_+) & a_+ & \dots & 0 \\
        \dots & \dots & \dots & \dots & \dots & \dots \\
        0 & 0 & 0 & 0 & 1 & 0 \\
        0 & 0 & 0 & 0 & 0 & 1 \\
        \end{array}
\right] . 
\]
Now we use with the following ansatz of the density profile, represented by
the row vector
\be
\langle \rho_{st}^{1,2}| =\{ \rho_-,\rho_b,\rho_+,\rho_b,...\rho_b \},
\label{conditionR1}
\ee
which must satisfy (as a property of the time-periodic steady state) 
\be 
\langle \rho_{st}^{2,3}| = \langle \rho_{st}^{1,2}| {\cal W}^{2, 3},
\ee
where the row vector $\langle \rho_{st}^{2,3}|$ can be  obtained by spatially 
translating (by one lattice unit) the density 
profile $\langle \rho_{st}^{1,2}|$:
\be 
\langle \rho_{st}^{2,3}| = \{ \rho_b,\rho_-,\rho_b,\rho_+, \rho_b,  \dots,
\rho_b \}.
\ee
Using the particle conservation
\be 
\rho_+ + \rho_- + (L-2) \rho_b = L \rho
\label{conservation_R1}
\ee 
and Eq. \ref{conditionR1}, we solve for the $\rho_{\pm}$ in terms of $a_{\pm}$
\begin{eqnarray}
\rho_+ = \frac{\rho}{1-a_+,} ~;~ \rho_- &= (1-a_-)\rho.
\label{rho+-R1}
\end{eqnarray}
The above equations give an exact profile of the traveling density wave, which
has a bump just in front of the rightmost defect site and has a trough separated
from the bump by two lattice spacings, i.e., a trough at the leftmost defect
site, as shown in Fig. \ref{x-rho-R1}. Note that, in the case of a single defect
studied previously in \cite{Chatterjee2014}, the bump was immediately followed by a trough, i.e., the gap between the bump and the trough was 
only one lattice spacing, as opposed to two lattice spacings here. As we 
show below, this qualitative difference gives rise to new interesting 
features like appearance of particle-hole symmetry in the current, albeit 
only  in the regime of large defect-velocity. 
%Due to this qualitative change in the structure of density profile, the particle
%current (see Eq. \ref{current-R1}) is actually different (compare with Eq. 20 in
%\cite{Chatterjee2014}).

\begin{figure}[h]
\begin{center}
\leavevmode
\includegraphics[width=9cm]{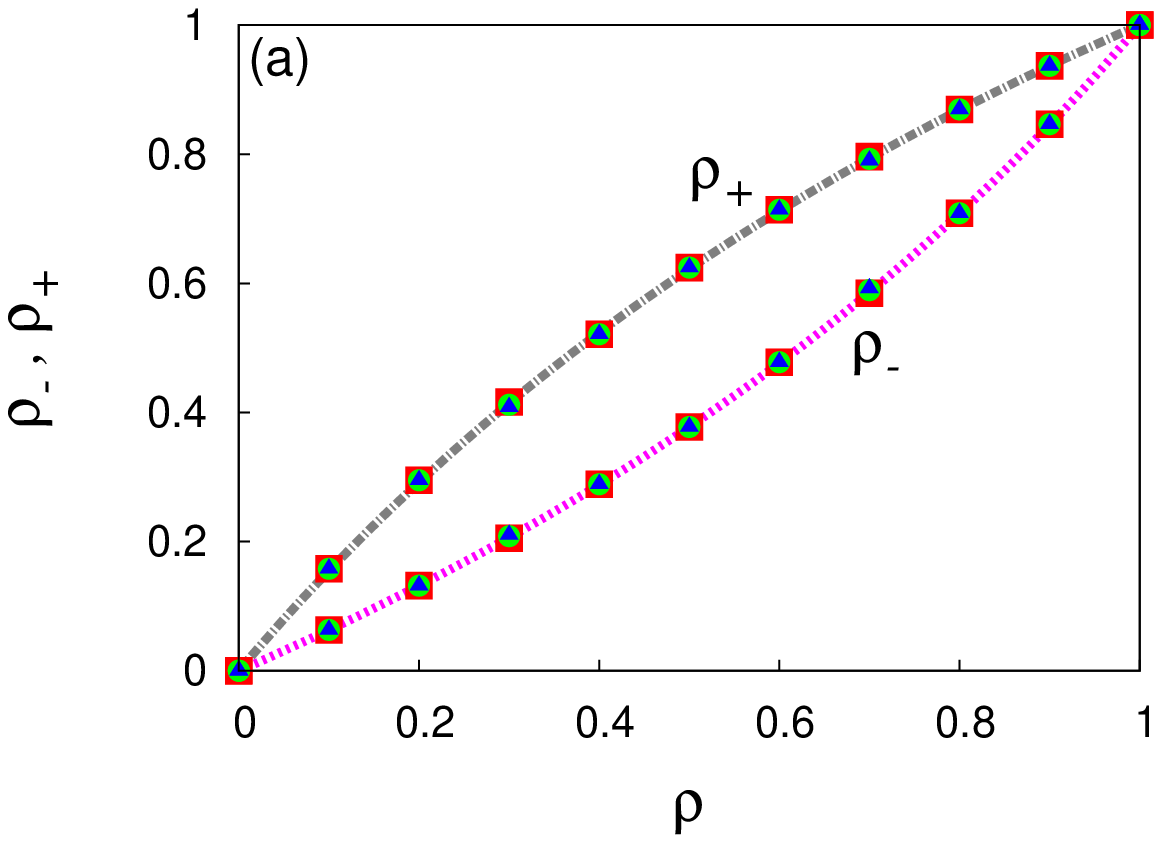}\includegraphics[width=9cm]{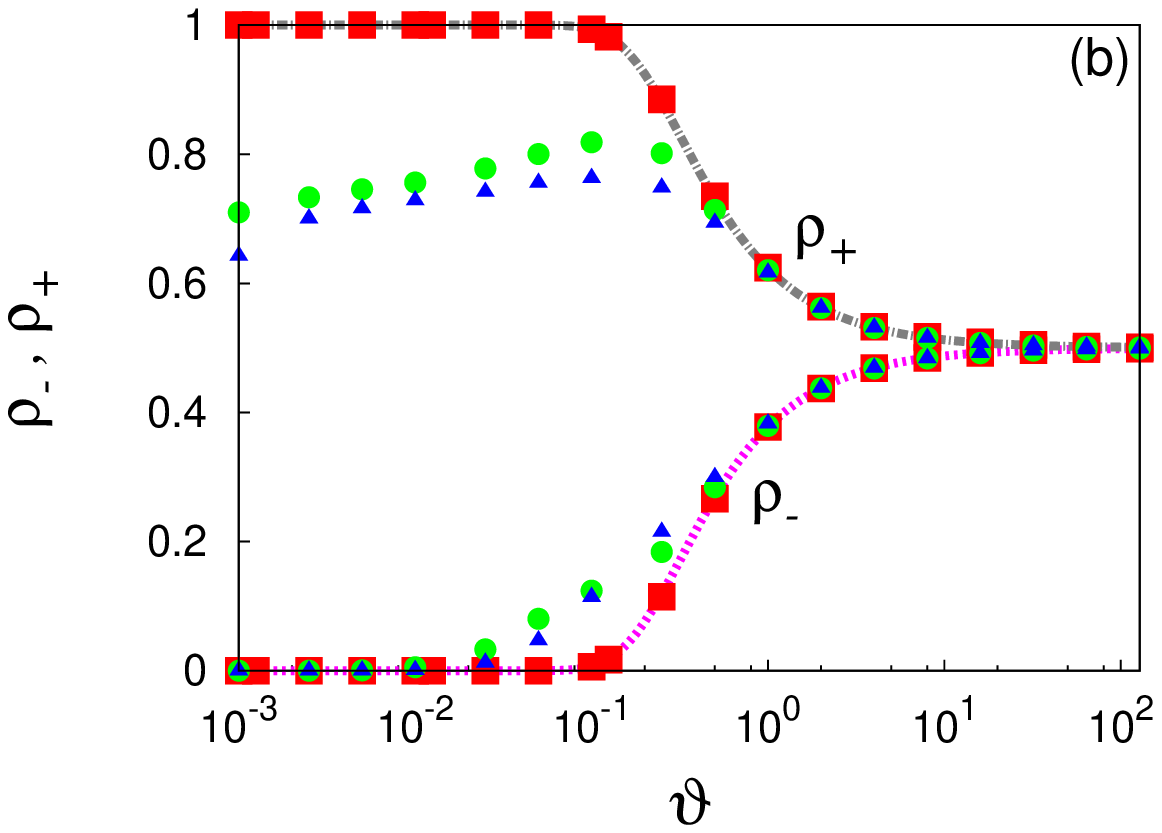}
\caption{(Color online) Two defects separated by distance $R=1$. (a) 
$\rho_+$ and $\rho_-$ are plotted against global density $\rho$ where defect 
velocity $v=1$ and $q=0$ (red squares), $ 0.2$ (green circles), $0.5$ (blue 
triangles). (b) $\rho_+$ and $\rho_-$ are plotted against defect 
velocity $v$ where $\rho=0.5$ and $q=0$ (red squares), $0.2$ (green circles),
$0.5$ (blue triangles).  In all the cases, $L=512$, $p=1$, $r=0$. Points
and lines are simulation and mean-field theoretical results, respectively. }
\label{rhopm-v-R1}
\end{center}
\end{figure}

Till now, the analysis remains exact. However, $a_{\pm}$ still contains
two-point density
 correlations (see Eq. \ref{a+-R1}) and needs to be calculated as a
function of the global density $\rho$. As mentioned before, solving the
hierarchy of equations involving many-point correlations is difficult and,
therefore, we resort to a mean-field method, where we approximate two-point
density correlations simply as a product of one-point correlations by writing 
$\langle \eta^{\alpha, \alpha+1}_{\alpha+2} \rangle \approx \rho_+$, 
$\langle \eta^{\alpha, \alpha+1}_{\alpha+2}
\eta^{\alpha, \alpha+1}_{\alpha + 3} \rangle \approx \rho_+ \rho_b$,
$\langle \eta^{\alpha, \alpha+1}_{\alpha + 1}\rangle \approx \rho_b$ and
$\langle \eta^{\alpha,\alpha+1}_{\alpha}
\eta^{\alpha,\alpha+1}_{\alpha+1} \rangle \approx \rho_- \rho_b$, to
obtain the conditional probabilities
\be 
a_+ = \frac{(1-\rho_b)}{\kappa} ~;~ a_-= \frac{(1-\rho_-)}{\kappa}.
\label{a+-R1_2}
\ee
Now we use Eqs. \ref{rho+-R1} and \ref{a+-R1_2} to eliminate $a_{\pm}$ and then
Eq. \ref{conservation_R1} to obtain $\rho_\pm$ as a function of global density
$\rho$ in the limit of large $L \gg 1$,
\bea
\rho_+ = \frac{\rho \kappa}{\rho+\kappa-1} ~;~ \rho_- = \frac{\rho(1-\kappa)}{\rho-\kappa}.
\label{rho+-R1_2}
\eea
Here, we have a rather simple closed form expression for $\rho_{\pm}$, unlike
 the case of a single defect in \cite{Chatterjee2014}.

In Fig. \ref{rhopm-v-R1}, we plot  $\rho_\pm$ as a function of the
global density $\rho$ and the defect velocity $v$, 
for $p=1$, $r=0$ and $q=0$ (red squares). The simulation results show good
agreement with our analytical expression in Eq. \ref{rho+-R1_2}. We also present
our numerical results for  $q \ne 0$ in the same plot. We find that for $v \gg
q$, $\rho_\pm$ do not depend strongly on $q$ and Eq. \ref{rho+-R1_2} remains
valid even for $q \ne 0$. However, for 
$v \lesssim q$ when the defect movement and bulk relaxation happen over similar
time-scales, nontrivial correlations develop in the system and the above mean-field
predictions break down, at least quantitatively; in Fig.
\ref{rhopm-v-R1}b, 
one can see that, in the regime $v \lesssim q$, the deviations
of the simulation results from the mean-field theory start creeping in. For very
small defect velocity $v \ll p, q$, the system approaches an
equilibrium state where the bump, expectedly, disappears (i.e., $\rho_+
\rightarrow \rho$) and the trough becomes completely devoid of any particle
(i.e., $\rho_- \rightarrow 0$).

\begin{figure}[h]
\begin{center}
\leavevmode
\includegraphics[width=8.8cm]{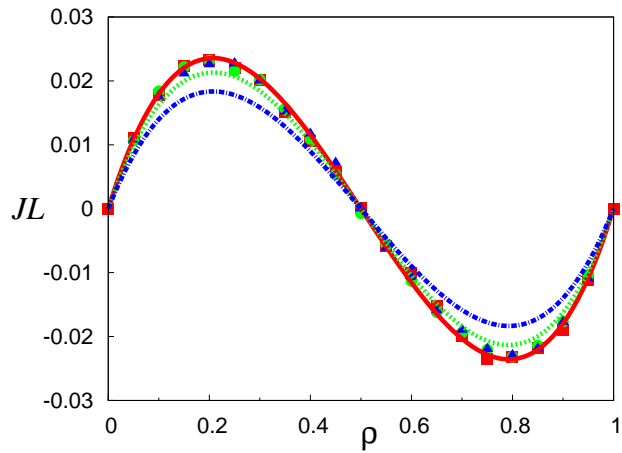}
\caption{(Color online) Two defects separated by distance $R=1$. Scaled current 
$JL$ is plotted against the global density $\rho$ with defect velocity $v=1.0$ 
and $L=512$, $p=1$, $r=0$ for various values of $q=0$ (red squares), $0.2$ 
(green circles), $0.5$ (blue triangles). The current is particle-hole symmetric 
and consequently vanishes at the half-filling $\rho=0.5$. }
\label{j-rho_R1}
\end{center}
\end{figure}

\begin{figure}[h]
\begin{center}
\leavevmode
\includegraphics[width=9cm]{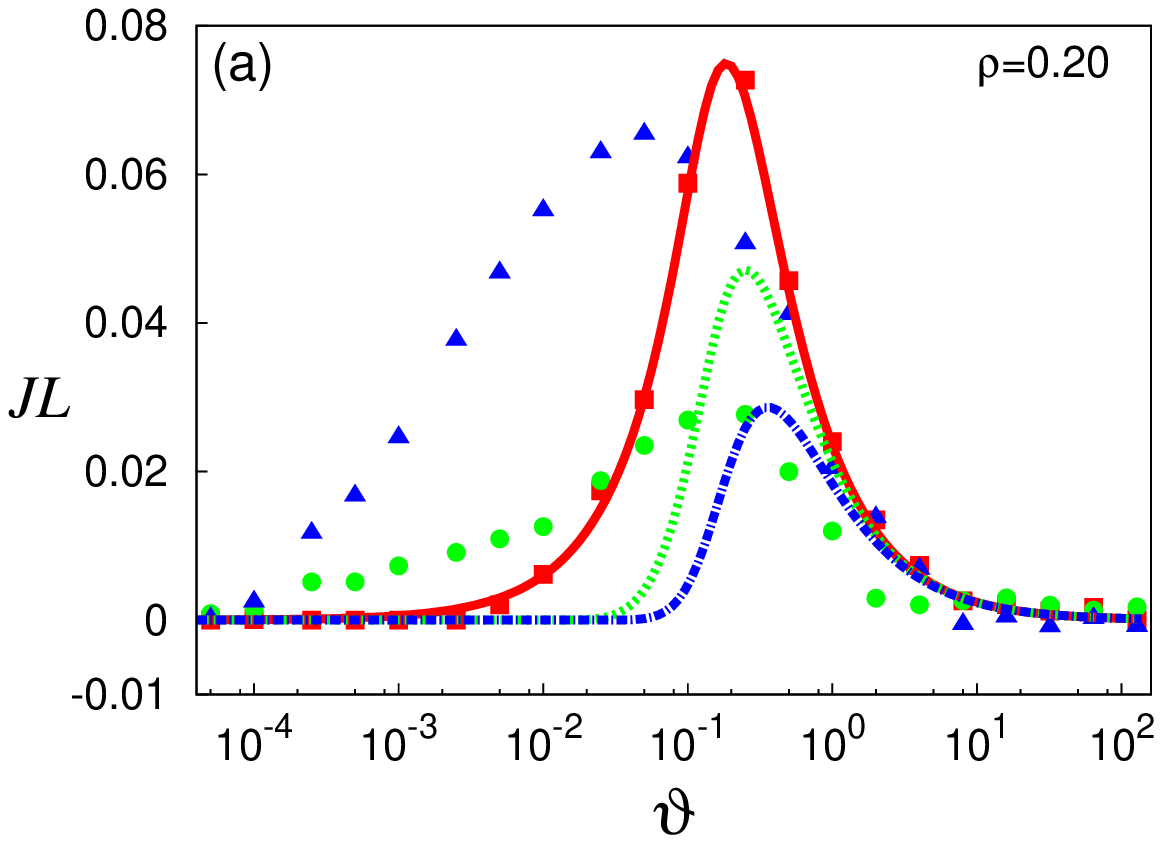}\includegraphics[width=9cm]{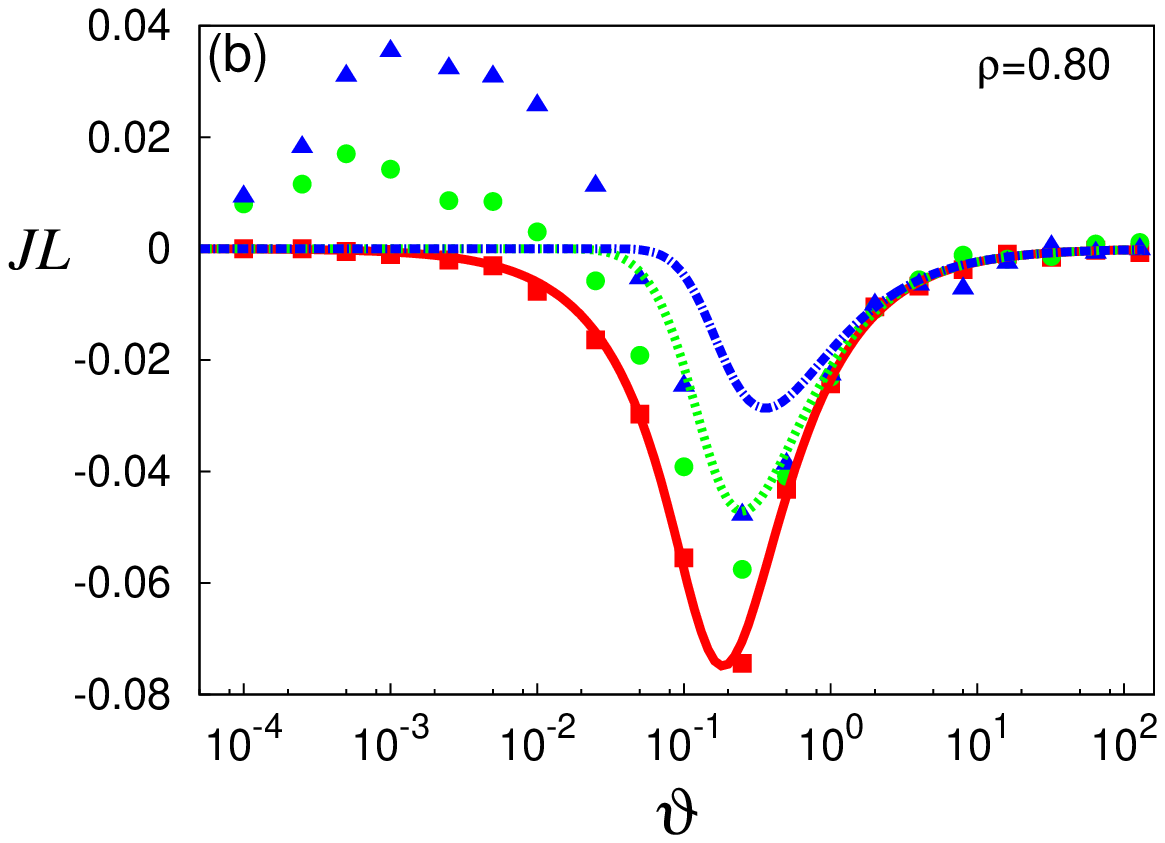}
\caption{(Color online) Two defects separated by distance $R=1$.  Scaled 
current $JL$ is plotted against defect velocity $v$ for two different
 values of densities (a) $\rho_1=0.2$ and (b) $\rho_2= 1- \rho_1 =0.8$. 
Although mean-field theory predicts a particle-hole symmetry, for $0 < v 
\lesssim q$, this breaks down. 
Here $q=0$ (red squares), $0.2$ (green circles), $0.5$ (blue triangles)
with $L=512$, $p=1$, $r=0$. The discrete points show simulation data and the 
continuous lines show mean-field predictions. }
\label{j-v_R1}
\end{center}
\end{figure}

Using the fact that Eq. \ref{rho+-R1_2} remains valid for all $q$ in the large
$v$ regime, one can calculate the current within mean-field theory for $v \gg
q$. It is easy to see that during the time-interval $\tau$, when the defect sites are
at $\alpha$ and $\alpha+1$, non-zero diffusive current exists only 
across four bonds in the system. Let $J_{i,i+1}$ denote the current across the
bond between the sites $i$ and $i+1$. Then the non-zero contributions to 
current can be written
as $J_{\alpha-1,\alpha} =
\tilde{q}[\rho(1-\rho_-)-\rho_-(1-\rho)]$, $J_{\alpha,\alpha+1} = -\tilde{p}\rho(1-\rho_-)$, $J_{\alpha+1, \alpha+2} = \tilde{q}[\rho(1-
\rho_+) - \rho_+ (1-\rho)]$ and
$J_{\alpha+2, \alpha+3} = \tilde{p} \rho_+ (1-\rho)$, 
where the effective Poisson hopping rates are 
$\tilde{p}=(v/L)(1-e^{-p/2v})$ and $\tilde{q}=(v/L)(1-e^{-q/2v})$. 
By adding all of the individual contribution to the current,
we obtain the net diffusive current averaged over a full cycle $L/v$, 
\bea
J_q(\rho,v) \simeq \tilde{q}[2\rho-\rho_--\rho_+] + \tilde{p}[\rho_+(1-\rho)-\rho(1-\rho_-)]. ~~~~
\label{current-R1}
\eea
The above mean-field expression of current differs not only quantitatively from
that in the single defect case (see Eq. 20 of Ref. \cite{Chatterjee2014}), 
but also qualitatively, because of
the particle-hole symmetry in the current where $J_q(\rho,v) = -J_q(1-\rho,v)$.
It can be easily checked from Eq. \ref{rho+-R1_2} that
 under the particle hole transformation $\rho
\rightarrow (1-\rho)$, the height and depth of the bump and the trough,
respectively, transform as follows: $\rho_+
\rightarrow 1-\rho_-$ and $\rho_- \rightarrow 1- \rho_+$. However, this holds
only for large $v$, since the mean-field expression of current in 
Eq. \ref{current-R1} breaks down for $0 < v \ll q$, where correlations play a 
nontrivial role.

\begin{figure}[h]
\begin{center}
\leavevmode
\includegraphics[width=8.8cm]{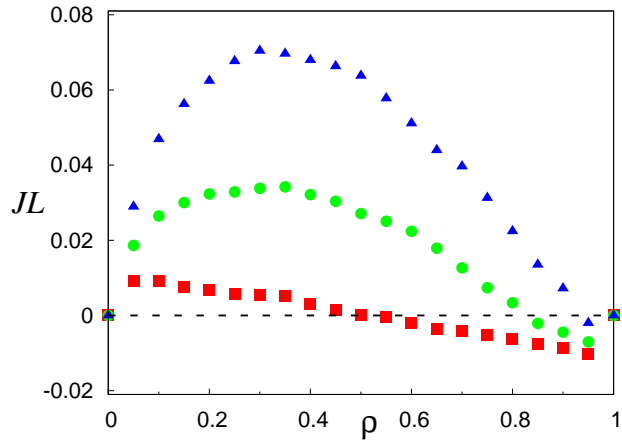}
\caption{(Color online) Two defects separated by distance $R=1$. Numerical 
simulation results of scaled current $JL$ is plotted as a function of the 
global density $\rho$ for $q=0$ (red squares), $0.2$ (green circles), $0.5$ 
(blue triangles)  where the defect velocity $v=10^{-3}$,
$L=512$ and $p=1$ and $r=0$. For intermediate
and small values of defect velocity $0 < v \lesssim q \ne 0$, the absence
of  particle-hole symmetry in the current is quite evident.}
\label{j-rho_sym-break}
\end{center}
\end{figure}

In Fig. \ref{j-rho_R1}, we plot scaled current $JL$ (scaled by system
size $L$) as a function of the global density $\rho$ for various values of $q,$
where defects move with velocity $v=1.0$. The current is particle-hole symmetric and
consequently vanishes for $\rho=0.5$. For
$q=0,$ the agreement between simulations and the mean-field theory as in Eq.
\ref{current-R1} is excellent. However,  some discrepancies between theory and
simulations are observed at larger values of $q$; though the particle-hole
symmetry is still found to be obeyed. In Fig. \ref{j-v_R1}, we have plotted
the scaled current $JL$ as a function of defect velocity $v$ for two different
densities $\rho_1 = 0.2$ and $\rho_2= 1-\rho_1 = 0.8$, 
which are related to each other by the particle-hole transformation. At
large velocities $v \gg q$, the mean-field theory 
captures quite well the broad features of the particle current. However, at the
intermediate regime of velocity $0 \ll v \ll q $, the mean-field theory
breaks down where the current, for  nonzero $q$, is actually found to
remain positive for both the densities $\rho_1=0.2$ and $\rho_2=0.8$, implying
that the  particle-hole symmetry is no more present at small values of $v$. Fig. 
\ref{j-rho_sym-break} highlights this point, where we have plotted the
scaled current as a function of density for a small value of $v=10^{-3}$.

Above analysis can be easily generalized for arbitrary number of defects located
at consecutive $n$ number of sites $\alpha_{k+1}=\alpha_{k}+1$ with $k=1, \dots,
(n-1)$. The ansatz for the density profile needs to be suitably modified but many 
of the qualitative conclusions including existence of particle-hole symmetry
remain valid.

\subsection{$R=2$: two next-nearest-neighbor defects with infinite potential barrier}

\begin{figure}[h]
\begin{center}
\leavevmode
\includegraphics[width=9cm]{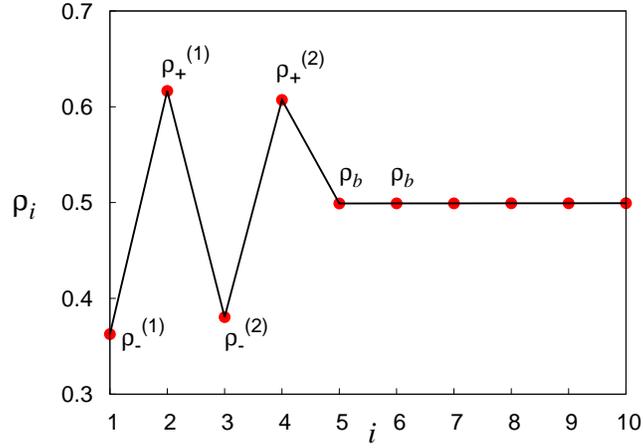}
\caption{(Color online) Two defects separated by distance $R=2$.  (here the 
defects reside at sites $1$ and $3$).  Density profile for defect sites $1$ and $3$
shows two peaks $\rho_+^{(1)},\rho_+^{(2)}$ and two troughs
$\rho_-^{(1)},\rho_-^{(2)}.$ Bulk density is denoted by $\rho_b$. }
\label{rhopm-v-R2}
\end{center}
\end{figure}

In this section, we analyze the case where the defects are separated by two
lattice spacings, i.e., if one defect is located at site $\alpha$, the other
one is located at site $\alpha+2$. We first consider the case with
$q=0$. Our numerical simulation results for the density profile are shown in
Fig. \ref{rhopm-v-R2} where two bumps and two troughs can be seen. 
Motivated by this, we use the following ansatz for the density profile,
measured at a time when the two defects are at sites $1$ and $3$, 
\be
\langle \rho_{st}^{1, 3}| =\left\lbrace \rho_-^{(1)},\rho_+^{(1)},
\rho_-^{(2)},\rho_+^{(2)},\rho_b....\rho_b \right\rbrace
\label{ansatz_R2}
\ee
where there are two bumps and two troughs, placed alternately on 
four adjacent sites: first a trough of density $\rho_-^{(1)}$ and, then on the three
right neighbours, a bump with density $\rho_+^{(1)}$, a trough with density
$\rho_-^{(2)}$ and a bump with density $\rho_+^{(2)}$. 

To study the density profile, we need to calculate the conditional probability
$a_+^{(1)} $ (or $a_-^{(1)}$) that, during the time-interval $\tau$, the
occupied defect site $\alpha$ exchanges particle with its right (left)
neighbour  and the conditional probability $a_+^{(2)}$ ($a_-^{(2)}$) that,
during the time-interval $\tau$, the occupied defect site  $\alpha+2$ exchanges
particle with its right (left) neighbour. For the case of $q = r = 0$, the
non-vanishing transition rates occur only for the particles which are hopping
from the defect sites, $\alpha$ and $\alpha+2$, for which the
transition matrix can be constructed in terms of these conditional probabilities
as given below,
\begin{widetext}
\[
{\cal W}^{2, 4}= \left[ \begin{array}{cccccccc}
1 & 0 & 0 & 0 & 0 & 0 & \dots& 0 \\
a_-^{(1)} & \left( 1-a_+^{(1)}-a_-^{(1)} \right) & a_+^{(1)} & 0 & 0 & 0 &\dots & 0 \\
0 & 0 & 1 & 0 & 0 & 0& \dots & 0 \\
0 & 0 & a_-^{(2)} & \left( 1-a_+^{(2)}-a_-^{(2)} \right) & a_+^{(2)} & 0 & \dots & 0 \\
0 & 0 & 0 & 0 & 1 & 0 & \dots & 0 \\
\dots & \dots & \dots & \dots & \dots & \dots & \dots & \dots \\
0 & 0 & 0 & 0 & 0 & \dots & 0 & 1 \\
        \end{array}
\right] \mbox{~~~~~~~~~~~~~~~~}
\]
\end{widetext}
In the time-periodic steady state, the following condition must be satisfied
\be
\langle \rho_{st}^{2, 4}| = \langle \rho_{st}^{1, 3}| 
{\cal W}^{2, 4} 
\label{condition_R2} 
\ee
where the $\langle \rho_{st}^{2, 4}|$ is given by
\be
\langle \rho_{st}^{2, 4}| =\left\lbrace \rho_b, \rho_-^{(1)},\rho_+^{(1)},
\rho_-^{(2)},\rho_+^{(2)},\rho_b....\rho_b \right\rbrace.
\ee
The steady-state condition Eq. \ref{condition_R2} leads to the following set of four equations,
\bea
\rho &=& \rho_-^{(1)} + \rho_+^{(1)} a_-^{(1)}, \\
\label{eq_set_mult_1}
\rho_-^{(1)} &=& \rho_+^{(1)}\left( 1-a_-^{(1)}-a_+^{(1)} \right), \\
\label{eq_set_mult_2}
\rho_-^{(2)} &=& \rho_+^{(2)}\left( 1-a_-^{(2)}-a_+^{(2)} \right), \\
\label{eq_set_mult_4}
\rho_+^{(2)} &=& \rho_+^{(2)} a_+^{(2)} + \rho,
\label{eq_set_mult_5}
\eea
along with the particle number conservation,
\be
\rho_-^{(1)} + \rho_+^{(1)} + \rho_-^{(2)} + \rho_+^{(2)} + (L-4)\rho_b = L\rho,
\label{conservation_R2}
\ee
which can be solved to exactly obtain $\rho_{\pm}^{(1)}$ and $\rho_{\pm}^{(2)}$ in terms
of the conditional probabilities $a_{\pm}^{(1)}$ and $a_{\pm}^{(2)}$,
\bea
\rho_+^{(1)} &=& \frac{\rho}{1-a_+^{(1)}}, 
\label{rho-R2-1} \\
\rho_-^{(1)} &=& \frac{\rho \left( 1-a_+^{(1)}-a_-^{(1)} \right)}{1-a_+^(1)}, 
\label{rho-R2-2} \\
\rho_+^{(2)} &=& \frac{\rho}{1-a_+^{(2)}}, 
\label{rho-R2-3} \\
\rho_-^{(2)} &=& \frac{\rho \left( 1-a_-^{(2)}-a_+^{(2)} \right)}{1-a_+^{(2)}}.
\label{rho-R2-4}
\eea

The conditional probabilities $a_{\pm}^{(1)}$ and $a_{\pm}^2$ can be written in
terms of three-point and four-point correlations as follows. First, we note that
a particle can hop only from one of the two defect sites $\alpha$ and
$\alpha+2$, either to the right or to the left, provided that the destination
site is empty. To calculate $a_+^{(1)}$, we need to consider rightward hopping 
of a particle from the defect site $\alpha$. For this hopping to take place, the
site $\alpha$ must be occupied and the site $\alpha +1$ must be empty. In
addition, it is crucial that within the residence time $\tau$, no leftward
hopping takes place from the other occupied defect site at $\alpha+2$ (as that
would block the site $\alpha +1$) and no leftward hopping takes place from the
defect site $\alpha$ (as the site $\alpha$ then gets empty). Therefore, in the
calculation of  $a_+^{(1)}$, the corresponding local configurations 
$1\hat{1}0\hat{0}$, $0\hat{1}0\hat{0}$,
$1\hat{1}0\hat{1}$ and $0\hat{1}0\hat{1}$ (``cap'' denotes the defect site, 
as before) are associated with different Poissonian decay rates
$1/\kappa_1 = (1-e^{-p/2v})$ and $1/2 \kappa_2 = (1 -e^{-p/v})/2$
\cite{Chatterjee2014} 
\begin{widetext}
\bea
a_+^{(1)} = \left[ \frac{\langle \eta^{\alpha, \alpha+2}_{\alpha} \eta^{\alpha, \alpha+2}_{\alpha+1}
(1 - \eta^{\alpha, \alpha+2}_{\alpha+2}) (1-\eta^{\alpha, \alpha+2}_{\alpha+3}) \rangle}{\kappa_1
\langle \eta^{\alpha, \alpha+2}_{\alpha+1} \rangle} \right] 
+ \left[ \frac{\langle (1 - \eta^{\alpha, \alpha+2}_{\alpha}) \eta^{\alpha, \alpha+2}_{\alpha+1}
(1-\eta^{\alpha, \alpha+2}_{\alpha + 2})(1 - \eta^{\alpha, \alpha+2}_{\alpha + 3})\rangle}{2 \kappa_2
\langle \eta^{\alpha, \alpha+2}_{\alpha+1} \rangle} \right] 
\nonumber \\
+ \left[ \frac{\langle \eta^{\alpha, \alpha+2}_{\alpha} \eta^{\alpha, 
\alpha+2}_{\alpha+1}(1-\eta^{\alpha, \alpha+2}_{\alpha+2}) \eta^{\alpha, \alpha+2}_{\alpha+3} 
\rangle}{2 \kappa_2 \langle \eta^{\alpha, \alpha+2}_{\alpha+1} \rangle} \right] 
+ \left[ \frac{\langle (1-\eta^{\alpha, \alpha+2}_{\alpha}) \eta^{\alpha, \alpha+2}_{\alpha+1}
(1-\eta^{\alpha, \alpha+2}_{\alpha+2}) \eta^{\alpha, \alpha+2}_{\alpha+3} \rangle}
{2 \kappa_2 \langle \eta^{\alpha, \alpha+2}_{\alpha+1} \rangle} \right]  
\label{eq_ap1}
\eea   
\end{widetext}
Similarly, we obtain the other conditional probabilities 
\begin{widetext}
\bea
a_-^{(1)} = \left[ \frac{\langle (1-\eta^{\alpha, \alpha+2}_{\alpha})
\eta^{\alpha, \alpha+2}_{\alpha+1} \eta^{\alpha, \alpha+2}_{\alpha+2} \rangle}{\kappa_1
\langle \eta^{\alpha, \alpha+2}_{\alpha+1} \rangle} \right] + \left[ \frac{\langle (1-\eta^{\alpha, \alpha+2}_{\alpha})
\eta^{\alpha, \alpha+2}_{\alpha+1}(1 - \eta^{\alpha, \alpha+2}_{\alpha+2}) \rangle}{2 \kappa_2
\langle \eta^{\alpha, \alpha+2}_{\alpha+1} \rangle} \right] 
\label{eq:a+-_1} 
\\
a_-^{(2)} = \left[ \frac{\langle (1-\eta^{\alpha, \alpha+2}_{\alpha+1})
(1 - \eta^{\alpha, \alpha+2}_{\alpha+2}) \eta^{\alpha, \alpha+2}_{\alpha+3} 
\eta^{\alpha, \alpha+2}_{\alpha+4} \rangle}{\kappa_1 \langle \eta^{\alpha, \alpha+2}_{\alpha+3} 
\rangle} \right] + \left[ \frac{\langle (1 - \eta^{\alpha, \alpha+2}_{\alpha+1}) 
(1 - \eta^{\alpha, \alpha+2}_{\alpha+2}) \eta^{\alpha, \alpha+2}_{\alpha+3}
(1 - \eta^{\alpha, \alpha+2}_{\alpha+4}) \rangle}{2 \kappa_2
\langle \eta^{\alpha, \alpha+2}_{\alpha+3} \rangle} \right] 
\nonumber \\
+ \left[ \frac{\langle \eta^{\alpha, \alpha+2}_{\alpha + 1} 
(1 - \eta^{\alpha, \alpha+2}_{\alpha+2}) \eta^{\alpha, \alpha+2}_{\alpha+3} 
\eta^{\alpha, \alpha+2}_{\alpha+4} \rangle}{2 \kappa_2 \langle \eta^{\alpha, \alpha+2}_{\alpha+3} \rangle} \right]
+ \left[ \frac{\langle \eta^{\alpha, \alpha+2}_{\alpha+1}
(1 - \eta^{\alpha, \alpha+2}_{\alpha+2}) \eta^{\alpha, \alpha+2}_{\alpha+3} 
(1 - \eta^{\alpha, \alpha+2}_{\alpha+4}) \rangle}{2 \kappa_2 \langle \eta^{\alpha, \alpha+2}_{\alpha+3} \rangle} \right]
\\
a_+^{(2)} = \left[ \frac{\langle \eta^{\alpha, \alpha+2}_{\alpha+2} \eta^{\alpha, \alpha+2}_{\alpha+3}
(1 - \eta^{\alpha, \alpha+2}_{\alpha + 4}) \rangle}{\kappa_1 \langle \eta^{\alpha, \alpha+2}_{\alpha+3} \rangle} \right] 
+ \left[ \frac{\langle (1 - \eta^{\alpha, \alpha+2}_{\alpha+2}) \eta^{\alpha, \alpha+2}_{\alpha+3} 
(1 - \eta^{\alpha, \alpha+2}_{\alpha + 4}) \rangle}{2 \kappa_2 
\langle \eta^{\alpha, \alpha+2}_{\alpha + 3} \rangle} \right]. 
\label{eq:a+-_2}
\eea
\end{widetext}
Now, using the mean-field approximations where we write the three- and
four-point correlations as a product of respective densities, we obtain the
following coupled set of four equations involving $a_{\pm}^{(1)}$, $a_{\pm}^{(2)}$,
$\rho_{\pm}^{(1)}$ and $\rho_{\pm}^{(2)},$
\begin{widetext}
\bea
a_+^{(1)} &=& \frac{\rho_-^{(1)}(1-\rho_-^{(2)})(1-\rho_+^{(2)})}{\kappa_1} + \frac{(1-\rho_-^{(1)})(1-\rho_-^{(2)})
(1-\rho_+^{(2)})}{2\kappa_2} 
+ \frac{\rho_-^{(1)}(1-\rho_-^{(2)})\rho_+^{(2)}}{2\kappa_2} + \frac{(1-\rho_-^{(1)})(1-\rho_-^{(2)})\rho_+^{(2)}}{2\kappa_2} 
\label{a-R2-1} \\
a_-^{(1)} &=& \frac{(1-\rho_-^{(1)})\rho_-^{(2)}}{\kappa_1} + \frac{(1-\rho_-^{(1)})(1-\rho_-^{(2)})}{2\kappa_2} 
\label{a-R2-2} \\
a_-^{(2)} &=& \frac{(1-\rho_+^{(1)})(1-\rho_-^{(2)})\rho}{\kappa_1} + \frac{(1-\rho_+^{(1)})(1-\rho_-^{(2)})(1-\rho)}{2\kappa_2} 
+ \frac{\rho_+^{(1)}(1-\rho_-^{(2)})\rho}{2\kappa_2} + \frac{\rho_+^{(1)}(1-\rho_-^{(2)})(1-\rho)}{2\kappa_2} 
\label{a-R2-3} \\
a_+^{(2)} &=& \frac{\rho_-^{(2)}(1-\rho)}{\kappa_1} + \frac{(1-\rho_-^{(2)})(1-\rho)}{2\kappa_2}.
\label{a-R2-4}
\eea
\end{widetext}
To solve the set of coupled Eqs. \ref{conservation_R2} - 
\ref{rho-R2-4} and Eqs. \ref{a-R2-1} - \ref{a-R2-4}, we use {\it Mathematica} and
obtain expressions for $\rho_\pm^{(1)}$ and $\rho_\pm^{(2)}$ as a function 
of global density $\rho$ and $v$.

\begin{figure}[h]
\begin{center}
\includegraphics[width=7.8cm]{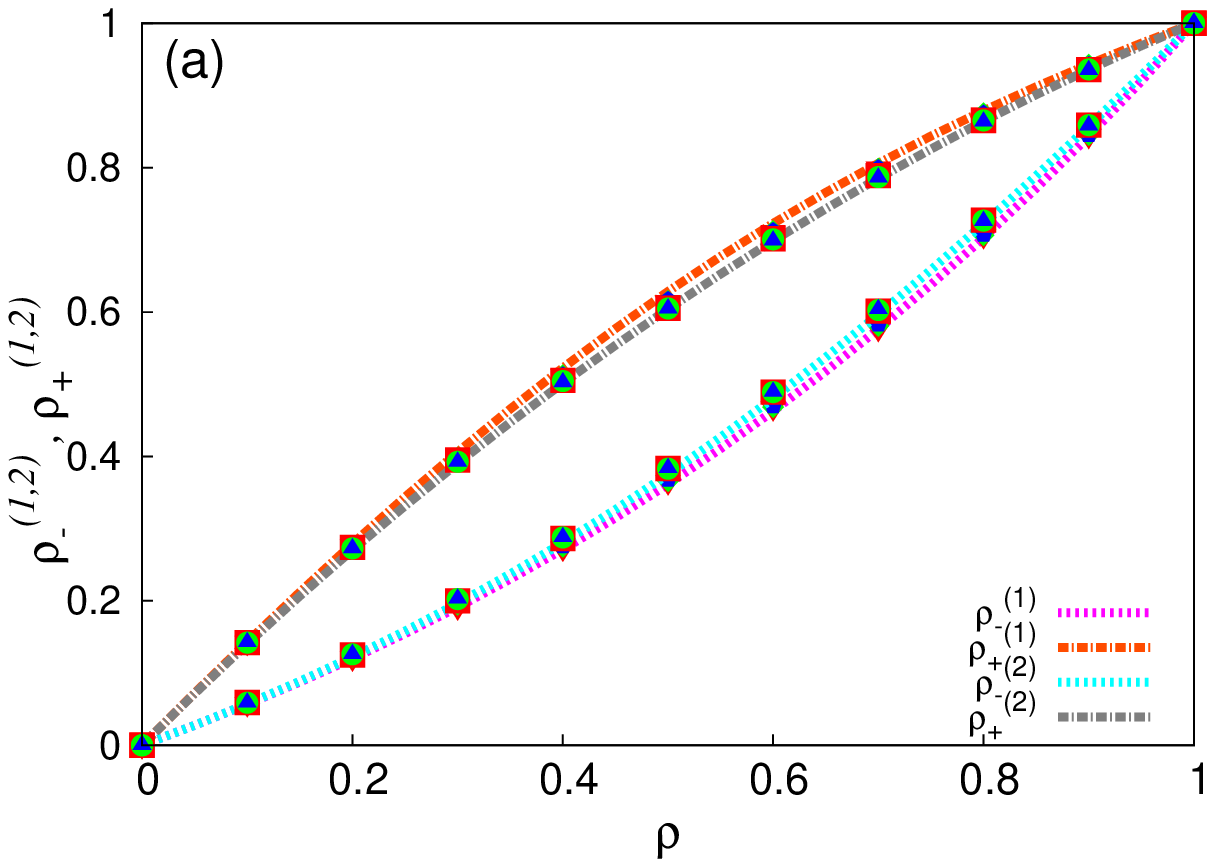}
\includegraphics[width=10.0cm]{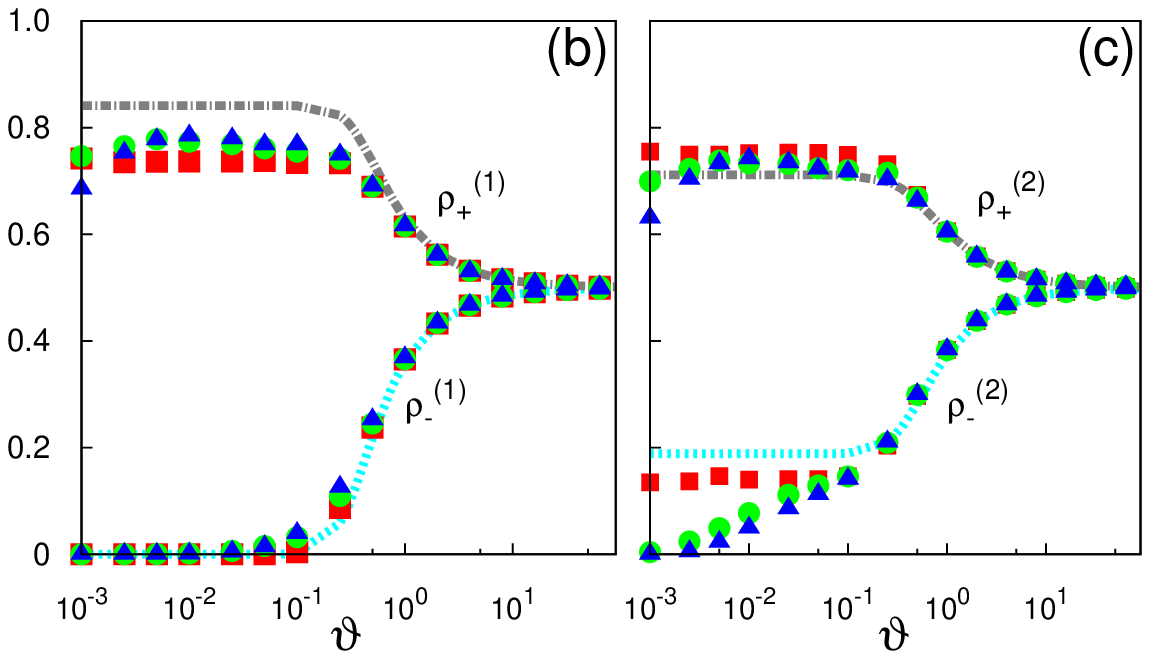}
\caption{(Color online) Two defects separated by
distance $R=2$. (a): Variation of $\rho_{\pm}^{(1)}$ and $\rho_{\pm}^{(2)}$
with global density $\rho$ for various values of $q$ with $v=1.0,$ $p=1,$ and 
$r=0$. (b) and (c): Variation of $\rho_{\pm}^{(1)}$ and 
$\rho_{\pm}^{(2)}$ with defect velocity $v$ for various 
$q$ with $\rho=0.5,$ $p=1$ and $r=0.$ Here, $q=0$ (red 
squares), $0.2$ (green circles) and $0.5$ (blue triangles). Lines are 
obtained from mean-field theory and discrete points from simulations.
}
\label{rhopm-rho-v-R2}
\end{center}
\end{figure}

In Fig. \ref{rhopm-rho-v-R2}, we plot $\rho_{\pm}^{(1)}$ and
$\rho_{\pm}^{(2)}$ as a function of global density $\rho$ (Fig. \ref{rhopm-rho-v-R2}a) and defect
velocity $v$ (Fig. \ref{rhopm-rho-v-R2}b and c) for various values of $q=0$, $0.2$ and $0.5$. For $q=0$, the
agreement between simulations and mean-field theory is excellent. For nonzero
$q$, the agreement is good at the large velocity regime $v \gg q$. However, some
discrepancies are observed for $v \lesssim q$. Also, note that the numerical
values of  $\rho_+^{(1)}$ and $\rho_+^{(2)}$ (or, $\rho_+^{(1)}$ and
$\rho_-^{(2)}$) are close but they are not exactly equal. 
Moreover, as we show below, unlike the cases of the two defects
with $R=1$ discussed in the previous subsection, there is no
particle-hole symmetry for two defects with $R=2$, even within mean-field
theory.

\begin{figure}[h]
\begin{center}
\includegraphics[width=9cm]{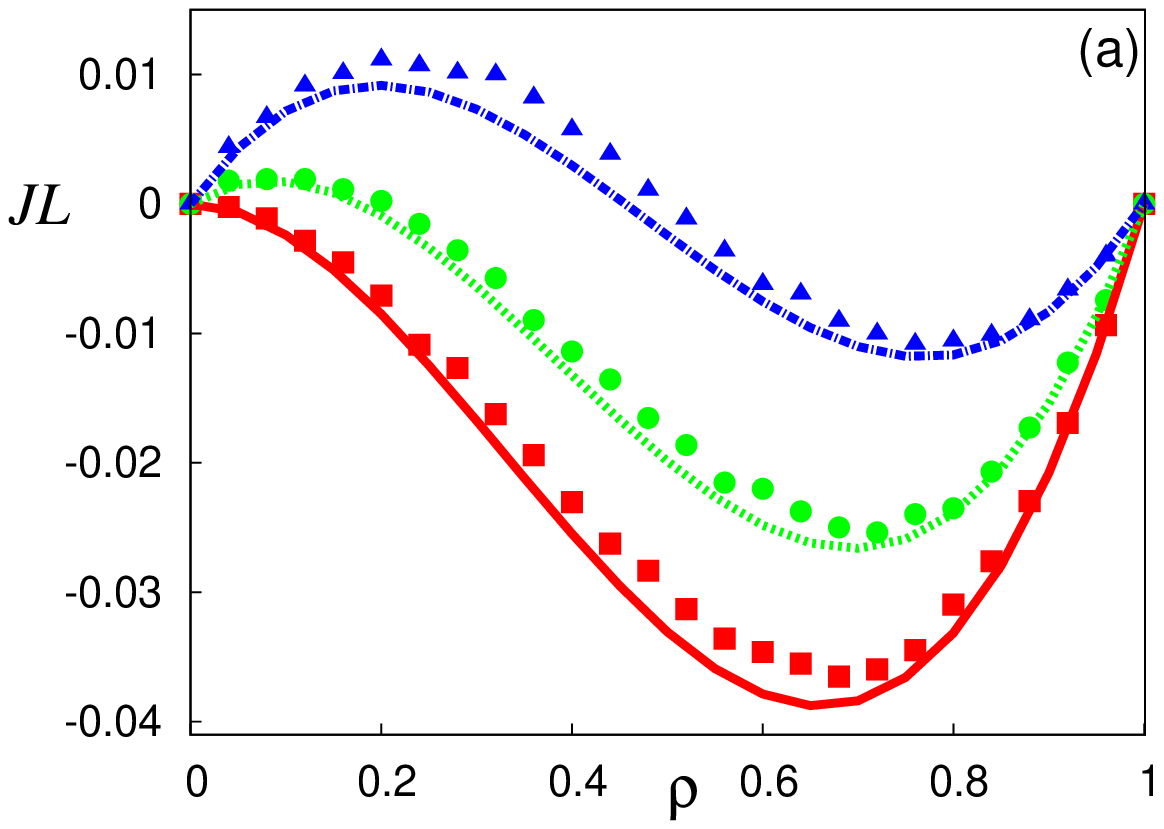}\includegraphics[width=9cm]{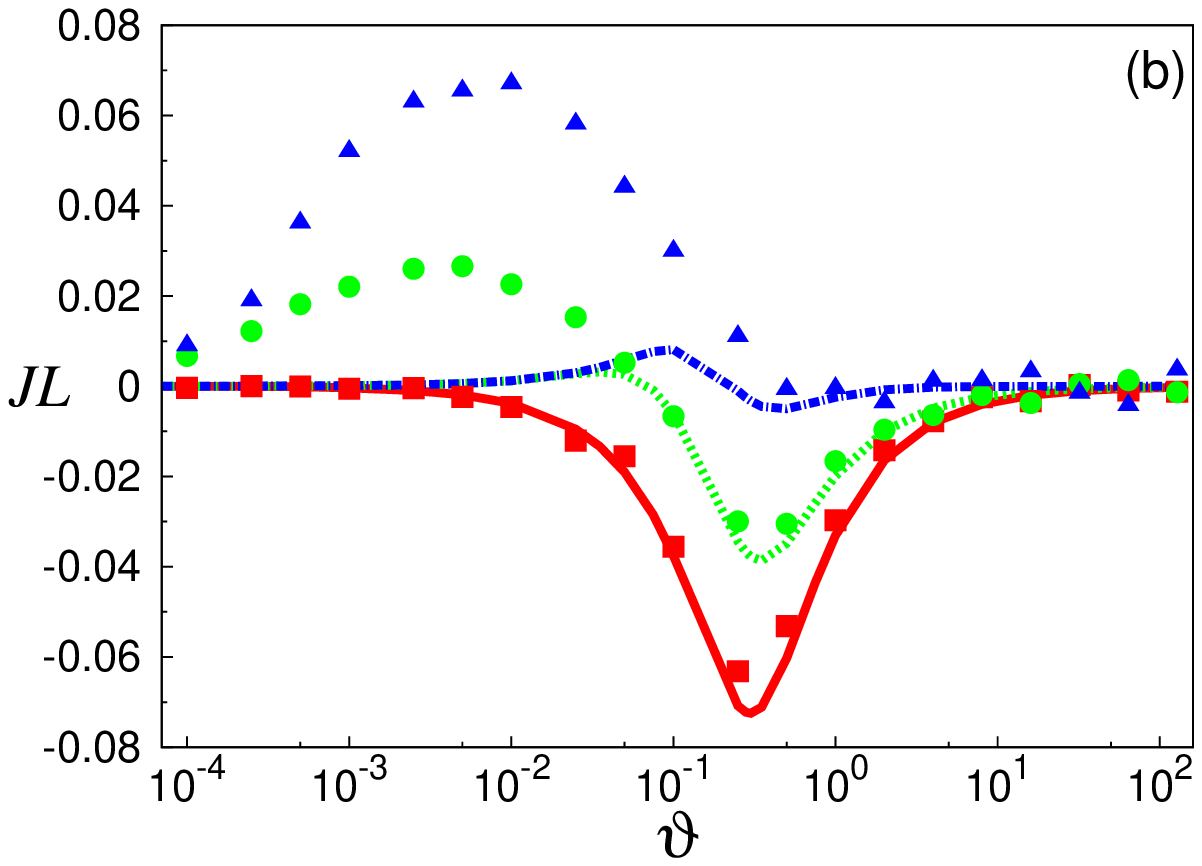}
\caption{(Color online )Two defects separated by distance $R=2$.  (a): 
Variation of the scaled current $JL$ with global density $\rho$ for  
$v=1$. (b): Variation of the scaled 
current $JL$ with defect velocity $v$ for $\rho=0.5$. Here, $q=0$ (red
squares), $0.2$ (green circles) and $0.5$ (blue triangles) and $p=1$ and $r=0$.
Solid lines are from mean-field theory and points from simulations.}
\label{current-rho-v-R2}
\end{center}
\end{figure}

To calculate the current within mean-field theory, we note that, for $q \neq 0$,
the following five bonds actually contribute to the current during the defect
residence time $\tau$. The individual diffusive currents, $J_{i,i+1}$ across the
bond between the sites $i$ and $i+1$, are given below,
\bea
J_{\alpha-1, \alpha} &=& \tilde{q}[\rho(1-\rho_-^{(1)}) - \rho_-^{(1)}(1-\rho)] =
\tilde{q} (\rho-\rho_-^{(1)}), \nonumber \\
J_{\alpha, \alpha+1} &=& -\tilde{p}[\rho_+^{(1)}(1-\rho_-^{(1)})], \nonumber \\
J_{\alpha+1, \alpha+2} &=& \tilde{p}[\rho_+^{(1)}(1-\rho_-^{(2)})], \nonumber \\
J_{\alpha+2, \alpha+3} &=& -\tilde{p}[\rho_+^{(2)}(1-\rho_-^{(2)})], \nonumber \\
J_{\alpha+3, \alpha+4} &=& \tilde{p}[\rho_+^{(2)}(1-\rho)] \nonumber
\eea
where the effective Poisson rates are $\tilde{p}=(v/L)(1-e^{-p/2v})$ and
$\tilde{q}=(v/L)(1-e^{-q/2v})$.
The net current can be obtained by adding all these individual contributions
\bea
J_q(\rho, v) = J_{\alpha-1, \alpha} + J_{\alpha, \alpha+1} 
+ J_{\alpha+1, \alpha+2} %\nonumber \\
+ J_{\alpha+2, \alpha+3} + J_{\alpha+3, \alpha+4}.
\eea
which we compare with the simulations in Fig. \ref{current-rho-v-R2}. The
agreement between simulations and mean-field theory is reasonably good for large
$v \gg q$. For intermediate velocity $v \lesssim q$, mean-field theory breaks
down due to nontrivial spatial correlations; however, it still captures the
broad features like current reversal as a function of defect velocity or
density.

\begin{figure}[h]
\begin{center}
\includegraphics[width=9cm]{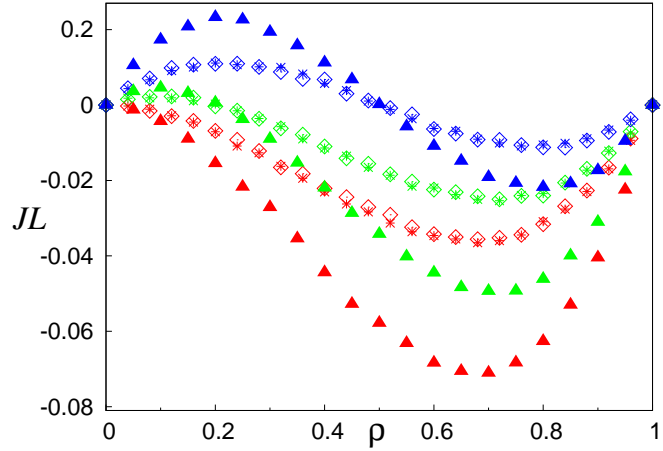}
\caption{(Color online) Comparison of scaled current $JL$ {\it vs.} density 
$\rho$ between three cases - (i) with a single defect (diamond; data from 
\cite{Chatterjee2014}) (ii) with two defects separated by distance $R=2$ 
(asterix) and (iii) $R=3,$ where defects reside in next nearest neighbors 
(triangle), for various values of $q=0$ (red), $0.2$ (green), and $0.5$ 
(blue); with $v=1$, $p=1$ and $r=0$.
Note that, for defect velocity $v \gtrsim q$ larger than bulk hopping rate, 
the two cases (i) and (ii) show approximately equal currents. 
For case (iii), the current is approximately doubled.}
\label{fig:rhoj_compare}
\end{center}
\end{figure}

\begin{figure}[h]
\begin{center}
\includegraphics[width=9cm]{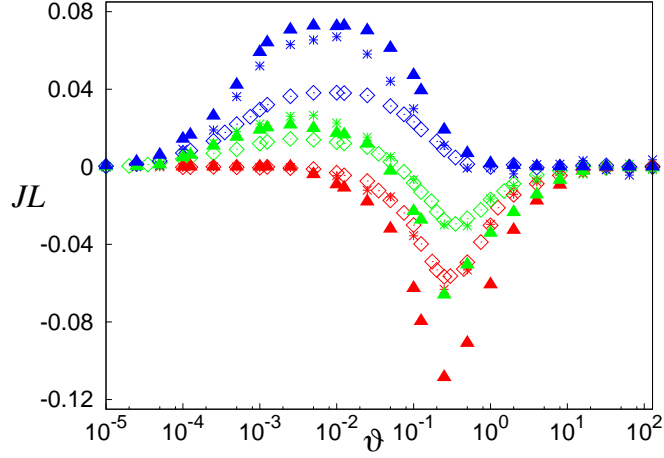}
\caption{(Color online) Comparison of scaled current $JL$ {\it vs.} velocity 
$v$ between three cases - (i) with a single defect (diamond), (ii) with two 
defects separated by distance $R=2$ (asterix) and (iii) $R=3,$ where defects 
reside at next nearest neighbors (triangle), for various values of $q=0$ 
(red), $0.2$ (green), and $0.5$ (blue); with $\rho=0.5$, $p=1$ and $r=0$ are 
fixed throughout. At larger velocity $v \gg q$, the currents in cases (i) and (ii) 
are approximately same. On the other hand, at smaller velocity $v \lesssim q$, the 
currents in cases (ii) and (iii) are quite close.}
\label{fig:vj_compare}
\end{center}
\end{figure}

Note that, if we assume $\rho_+^{(1)} \approx \rho_+^{(2)} = \rho_+$ (which is
indeed the case as seen in Fig. \ref{rhopm-rho-v-R2}), 
two currents $J_{\alpha+1, \alpha+2}$ 
and $J_{\alpha+2, \alpha+3}$ cancel each other. Further assuming
$\rho_-^{(1)} \approx \rho_-^{(2)}= \rho_-$, we obtain, to a good approximation,
the net current 
\be 
J_q(\rho, v) \simeq \tilde p \rho_+(\rho_- - \rho) + \tilde q (\rho_- - \rho),
\label{J-R2}
\ee 
which matches the expression for net current derived in the case of a single
moving defect in \cite{Chatterjee2014}. We verify this in Figs
\ref{fig:rhoj_compare} and \ref{fig:vj_compare}, where we plot the scaled
current $JL$ against $\rho$ and $v$, respectively,
 and compare with our data in the single
defect case. We find for large $v \ge q$, the current is indeed same in the two cases
but for smaller $v \ll q$, there is a significant difference.
%Interestingly, as we discuss below in the next section, the current in the 
%case of $R \ge 3$ becomes almost double of that in the case of $R=2$.
 The above analysis can be extended 
to arbitrary $n$ number of defects located at alternate sites 
$\alpha_{k+1} = \alpha_k+2$ with $k=1, \dots, (n-1)$.

\subsection{$R \ge 3$: Two defects separated by three or more lattice spacing}

In this section, we consider the case when the defect sites are separated by three 
or more lattice spacing $R \ge 3$. We present explicit results for $R=3$, and these
results can be easily extended to the cases with $R>3$.  Let us assume the defect
sites are located at $\alpha$ and $\alpha +3$. Recall that the density
pattern created by a single defect consists of a trough at the defect site
and a bump at the next site. Therefore, when the two defects are at a distance
$R=3$, the density pattern created around each of them do not overlap and as a
result, the local density at site $\alpha+2$ remains same as the bulk density.
From this simple consideration, we formulate the ansatz,
\be
\langle \rho_{st}^{1, 4}| =\left\lbrace \rho_-^{(1)},\rho_+^{(1)}, \rho_b, \rho_-^{(2)},
\rho_+^{(2)},\rho_b....\rho_b \right\rbrace
\label{ansatz_R3},
\ee
where, at the time of measurement, the defects are at sites $1$ and
$4$, and are about to move on to sites $2$ and $5$. 
The corresponding transition matrix has the form 
\begin{widetext}
\[
{\cal W}^{2, 5}= \left[ \begin{array}{ccccccccc}
1 & 0 & 0 & 0 & 0 & 0 & 0 & \dots & 0 \\ 
 a_-^{(1)} & \left( 1-a_+^{(1)}-a_-^{(1)} \right) & a_+^{(1)} & 0 & 0 & 0 & 0 &
\dots & 0 \\
0 & 0 & 1 & 0 & 0 & 0 & 0 & \dots & 0 \\
0 & 0 & 0 & 1 & 0 & 0 & 0 & \dots & 0 \\
0 & 0 & 0 & a_-^{(2)} & \left( 1-a_+^{(2)}-a_-^{(2)} \right) & a_+^{(2)} & 0 & \dots & 0 \\
0 & 0 & 0 & 0 & 0 & 1 & 0 & \dots & 0 \\
\dots & \dots & \dots & \dots & \dots & \dots & \dots & \dots & \dots \\
0 & 0 & 0 & 0 & 0 & 0 & \dots & 0 & 1 \\
\end{array}
\right] \mbox{~~~~~~~~~~~~~~~~}
\]
\end{widetext}
which acts on the above state vector  $\langle \rho_{st}^{1, 4}|$ to give the
new state vector $\langle \rho_{st}^{2, 5}| = \{ \rho_b,
\rho_-^{(1)},\rho_+^{(1)}, \rho_b, \rho_-^{(2)},\rho_+^{(2)},\rho_b....\rho_b
\}$, i.e.,
\be 
\langle \rho_{st}^{2, 5}| = \langle \rho_{st}^{1, 4}| {\cal W}^{2, 5}
\label{condition_R3}.
\ee
The steady-state condition Eq. \ref{condition_R3} yields exactly
the same relations, between density peaks, troughs and the conditional
probabilities, as given in Eqs. \ref{rho-R2-1} - \ref{rho-R2-4}. The conditional
probabilities $a_{\pm}^{(1)}$ and $a_{\pm}^{(2)}$, within mean-field
approximation, can be written as
\bea
a_+^{(1)} =  \frac{(1-\rho)\rho_-^{(1)}}{\kappa_1} + \frac{(1-\rho)(1-\rho_-^{(1)})}{2\kappa_2},
\\
a_-^{(1)} = \frac{(1-\rho_-^{(1)})\rho}{\kappa_1} + \frac{(1-\rho)(1-\rho_-^{(1)})}{2\kappa_2},
\label{apm-1-R3}
\eea
and 
\bea
a_+^{(2)} =  \frac{(1-\rho)\rho_-^{(2)}}{\kappa_1} + \frac{(1-\rho)(1-\rho_-^{(2)})}{2\kappa_2},
\\
a_-^{(2)} = \frac{(1-\rho_-^{(2)})\rho}{\kappa_1} + \frac{(1-\rho)(1-\rho_-^{(2)})}{2\kappa_2}
\label{apm-2-R3}
\eea
which are however different from Eqs. \ref{a-R2-1} - \ref{a-R2-4}. Note
that the two sets of equations - one involving $\rho_{\pm}^{(1)}$ and
$a_{\pm}^{(1)}$  (Eqs \ref{rho-R2-1}, \ref{rho-R2-2} and \ref{apm-1-R3}) and the
other involving $\rho_{\pm}^{(2)}$ and $a_{\pm}^{(2)}$ (Eqs \ref{rho-R2-3},
\ref{rho-R2-4} and \ref{apm-2-R3})- are decoupled from each other and imply
$\rho_{\pm}^{(1)} = \rho_{\pm}^{(2)} = \rho_{\pm}$. The above analysis can be
straightforwardly extended to arbitrary number of defects, which are separated
from each other by three or more lattice spacings. Therefore, on the mean-field
level, two (or more) defects separated by distance $R \ge 3$ do not have any
effect on each other and act as collection of two (or more) isolated defects.
Consequently, the time-averaged current can be written as sum of the two
contributions arising from each of the defects, 
\be 
J_q(\rho, v) = 2 [\tilde p \rho_+(\rho_- - \rho) + \tilde q (\rho_- - \rho)],
\label{J-R3}
\ee 
which may be compared with Eq. \ref{J-R2}, the current in the case of $R=2$.

\section{SSEP with ordered sequential update}
\label{sec-4}

In this section, we show that a close connection exists between our system 
of SSEP with a single moving defect and an SSEP with ordered (sitewise) 
sequential update. In the latter process, $N$ hardcore (otherwise 
non-interacting) particles are considered on a ring consisting of $L$ 
sites and, in each Monte Carlo step, the $L$ sites of the
lattice are updated sequentially in a particular order. For example, 
site $1$ is chosen first and if there is a particle present on that site, 
it hops to any of the nearest neighbor sites with equal probability, 
provided the hardcore exclusion is satisfied. Then site $2$ is chosen and 
the same process is repeated. After that site $3$ is chosen and so on. 
This way all the sites on the lattice are accessed sequentially and 
updated. When the
$L$-th update is performed on the $L$-th site, that concludes one Monte Carlo
step.   

Note that the sequential movement of
the update-site is very similar to the movement of a single defect site in 
the limit $q=0$, $r=0$ and $v=L$. In other words, one could think of an 
infinite potential barrier present at the position of a single defect, which 
moves one lattice unit in every micro time-step and consequently performs 
one complete cycle every Monte Carlo step.   
Since no bulk hopping is allowed, in this case, hopping can take
place only at the defect site, no where else on the lattice. This model of SSEP with a single defect is just like the
case of the SSEP with sitewise ordered sequential update, where particles can only hop out of the site, which is being updated at a particular time-step. 
As we show below, because of this connection, the structure of
the density profile and the behavior of current in SSEP with sitewise ordered
sequential update is  very similar to what we find in the moving defect problem. 
There is, however, one subtle difference between these two models. In the 
moving-defect problem, the update sites are still chosen randomly and, when the defect site does not get selected, the occupancy of the system does not change. 
This is clearly different from
sitewise ordered sequential update. As shown later, due to this important
difference, strong correlations are observed for sitewise ordered 
sequential update, whereas mean-field theory is known to work rather well 
for the moving-defect problem with $q=0$ and large $v$.

\begin{figure}[h]
\includegraphics[scale=0.7,angle=0]{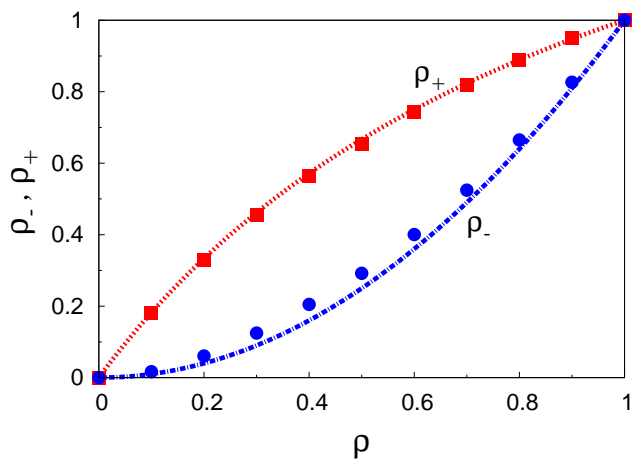}
\caption{(Color online) Density values $\rho_+$ and $\rho_-$ at the peak and the trough are
plotted as a function of $\rho$ for ordered (sitewise) sequential
update; system size $L=128$. Solid lines and points are obtained from mean-field 
calculation and simulation, respectively.}
\label{fig:ord}
\end{figure}

\begin{figure}
\begin{center}
\leavevmode
\includegraphics[width=9.0cm,angle=0]{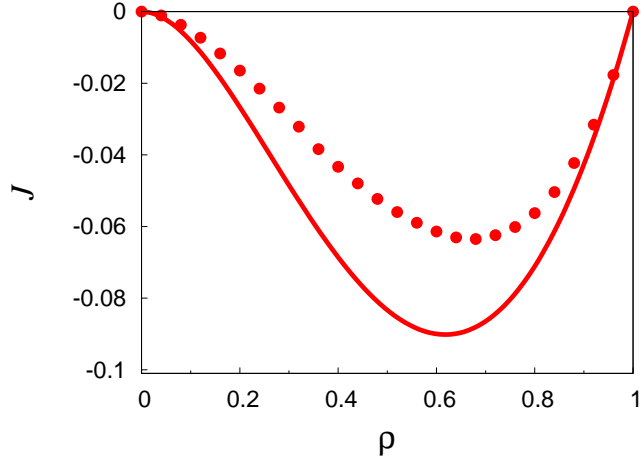}
\caption{(Color online) Current is plotted as a function of density $\rho$ for 
ordered (sitewise) sequential update with system size $L=128.$
Solid line is obtained from the mean-field theory and points from numerical 
simulation. The reason for the discrepancy
between mean-field theory and the simulation is due to the presence of
nearest-neighbour correlations around the site where the update is done.}
\label{j_vs_rho2}
\end{center}
%\vskip-6mm
\end{figure}

Time evolution equation for density in the case of sitewise ordered 
sequential updates can be written as 
\be
<\rho^{(\alpha+1)}(t_{micro}+1)| = <\rho^{(\alpha)}(t_{micro})| {\cal W}^{(\alpha+1)} 
\ee 
where matrix ${\cal W}^{(\alpha)}$ is the time evolution operator, $i$th element
of the row vector $<\rho^{(\alpha)}(t_{micro})|$ represents the density at site
$i$, $\alpha$ denotes the site which is being updated at a particular micro-time
$t_{micro}$. The density is measured at the end of the update process at
$t_{micro}$. The time evolution matrix ${\cal W}^{(\alpha)}$ can be
straightforwardly constructed as in the previous cases of the SSEP with moving
defects. For example, for specific values of $\alpha =1$ and $2$, ${\cal
W}^{(\alpha)}$ can be explicitly written as given below 
\[
{\cal W}^{(1)}= \left[ \begin{array}{cccccc}
        (1-a_+  - a_-) & a_+ & 0 & \dots & 0 & a_- \\
        0 & 1 & 0 & 0 & \dots & 0 \\
        \dots & \dots & \dots & \dots & \dots & \dots \\
        \dots & \dots & \dots & \dots & \dots & \dots \\
        0 & \dots & 0 & 0 & 1 & 0 \\
        0 & 0 & \dots & 0 & 0 & 1 \\
        \end{array}
\right], \mbox{~~~~~~~~~~~~~~~~}
\]

\[
{\cal W}^{(2)}= \left[ \begin{array}{cccccc}
        1 & 0 & 0 & \dots & 0 & 0 \\
        a_- & (1-a_+-a_-) & a_+ & 0 & \dots & 0 \\
        0 & 0 & 1 & 0 & \dots & 0 \\
        \dots & \dots & \dots & \dots & \dots & \dots \\
        0 & \dots & 0 & 0 & 1 & 0 \\
        0 & 0 & \dots & 0 & 0 & 1 \\
        \end{array},
\right], \mbox{~~~~~~~~~~~~~~~~}
\]
where the conditional probabilities 
\bea a_+ &=& \frac{\rm{Prob}[\eta^{(\alpha)}_{\alpha+2} = 0|\eta^{(\alpha)}_{\alpha+1} = 1]}{2},
\\
a_-& =& \frac{\rm{Prob}[\eta^{(\alpha)}_{\alpha} = 0|\eta^{(\alpha)}_{\alpha+1} =
1]}{2},
\eea
can be calculated in terms of nearest neighbours correlations.  
Time-periodic steady-state density profile satisfies the
following condition 
\be <\rho_{st}^{(\alpha+1)}| =
<\rho_{st}^{(\alpha)}| {\cal W}^{(\alpha+1)}. 
\label{condition_OSU}
\ee 
Without any loss of generality, we now consider $\alpha=1$
and proceed with an ansatz for $<\rho_{st}^{(1)}| = \{ \rho_{_-}, \rho_{_+},
\rho_{_b} \dots, \rho_{_b}, \rho_{_b} \}$. From the condition Eq.
\ref{condition_OSU}, we have $<\rho_{st}^{(2)}| = <\rho_{st}^{(1)}| {\cal
W}^{(2)}$, leading to
\bea \rho_{_+} &=& \frac{1}{1-a_+} \rho_{_b} > \rho_{_b}, \\
\rho_{_-}& =& \frac{1-a_+-a_-}{1-a_+} \rho_{_b} < \rho_{_b},
\\ 
\rho_{_b}& =& \frac{(1-a_+)L}{2-a_+-a_-+(1-a_+)(L-2)} \rho \simeq \rho, \eea
in the limit of large $L \gg 1$. Clearly, the density profile has the same
spatial structure having a density peak and trough, moving with a velocity $L$
per unit MC time. Time averaged particle current can be written as \be
J = \frac{1}{2} \left[ \langle \eta^{(\alpha)}_{\alpha+1} (1-\eta^{(\alpha)}_{\alpha+2}) \rangle  -
\langle \eta^{(\alpha)}_{\alpha+1} (1-\eta^{(\alpha)}_{\alpha}) \rangle \right]. 
\ee 
 Resorting to mean-field approximations, we get the following simple expressions,
\bea 
\rho_{_+} &=& \frac{2 \rho}{1 + \rho},
\\ 
\rho_{_-} &=& \rho^2
\\ 
J &=& - \frac{\rho^2 (1-\rho)}{1+\rho} < 0. 
\eea 
Here, the current is always opposite to the direction in which the ordered 
sequential updates are done and is {\it not} particle-hole symmetric. In 
Figs. \ref{fig:ord} and \ref{j_vs_rho2} we compare the mean
field predictions with simulation results. We find that, due to the 
nontrivial spatial correlations, the mean-field approximation does not work well here.

Previously, in Ref. \cite{Rajewsky}, a {\it bondwise} ordered sequential update was studied where
one after another bond (instead of sites) was chosen and updated. This particular update
process however, satisfies detailed balance and yields the same 
steady-state weights as in an ordinary SSEP with usual 
random sequential update, i.e.,  uniform 
measure in the configuration space. Thus, the SSEP with bondwise ordered sequential update
has a uniform density profile and does not have any current in the steady
state. The sitewise ordered sequential update, on the other hand, shows a
time-periodic steady state and a nonzero current as shown above.

\section{Conclusions}
\label{sec-5}

In this paper, we have studied the effect of time-periodic drive on a 
system of hardcore particles on a ring. We have modelled the effect of 
the drive by considering defects that move periodically on the lattice with 
uniform velocity. The particles diffuse on the lattice where the particle 
hopping rates, otherwise symmetric and uniform, are modified at and around the 
instantaneous position of the defects. Depending on these modified rates, the 
defects act like a moving potential barrier or well. We have demonstrated 
that, in certain limits, this model can be mapped onto a simple symmetric 
exclusion process with sitewise ordered sequential updates.

Here, we have explored the collective effects of the moving defects on the 
spatial structure and transport in the system. When defects are far apart, they effectively act like a collection 
of single individual defects. However, when the defects are close, spatial 
structures may be quite complex, e.g., multiple peaks and 
troughs could develop in traveling density wave. Consequently, the particle
current is described by fourth or higher order spatial correlations. 
In particular, when the defects occupy nearest neighbour positions on the 
lattice, in the limit of large defect velocity, the current shows 
particle-hole symmetry, which is not seen in any other cases. In general,
the particle current shows polarity reversal and non-monotonicity upon variation of particle density and defect velocity.

For simplicity, here we have considered an infinite potential 
barrier, i. e., $r=0$, which strictly forbids any particle to hop into the 
defect sites. A finite barrier would mean $r \neq 0$. In that case, we have 
verified (data not shown here) that the qualitative behaviors of the density 
profile and the current do not change. A nonzero $r$ merely tends to 
homogenize the density variations around the defect sites by reducing the 
height (depth) of the density peaks (troughs). Also, throughout the paper, we 
have considered defects which act like potential barrier, i.e., $p > r$. One 
may ask what happens for $p < r$, i.e., when a potential well is present 
instead. However, since a particle hopping out of a defect site (or into 
the defect site) with rate $p$ (rate $r$) can equivalently be described as 
a hole hopping into (or out of) the defect site with the same rate, 
the density profile and the current for the moving potential well can be 
obtained simply by substituting $\rho \rightarrow 1-\rho$ (interchanging the 
particles and holes) and $J \rightarrow -J$ in the respective quantities in 
the case of the barrier.

Our analytical formalism exploits the time-periodic nature
of the steady states, where we perform all measurements at
discrete time instants (stroboscopic measurements), when the
defects are about to leave a site and move onto the next one. We
represent the density profile as a traveling wave that has a local
inhomogeneity, at and around the defect site, and is uniform in
the bulk. To calculate the density inhomogeneities (peaks and
troughs in the density profile), we employ a mean-field theory.
At this point, it may be pertinent to make a few remarks about
another technique, viz. domain wall theory (DWT), which is
often used to obtain spatial steady-state structure in driven
diffusive systems \cite{kol2, dudz2, santen2, popkov2}. 
In the DWT, it is assumed that a
macroscopic domain wall forms in the system that separates
two spatially uniform domains with different densities. The
stochastic dynamics of the domain wall is described in terms
of a random walker whose position in the long time limit
decides the spatial structure of the density in the system.
In our system, the density inhomogeneity is present at and
around the defect site; i.e., the inhomogeneity is local, and
whenever the defect site moves, the inhomogeneous structure
also moves along with the defect. Within the stroboscopic
measurement, the movement of this density pattern is therefore
deterministic. To study the dissipation of a pattern at a certain
position and its formation at a new position, one has to
go beyond the stroboscopic frame and monitor the density
variation in continuous time. It may be of interest to consider
if a modified version of the DWT could be applied in that
case.

In all the cases discussed in this paper, the striking features in particle 
transport such as polarity reversal and appearance of multiple peaks in 
particle current as a function of defect velocity and density, remarkably 
persist, irrespective of the microscopic details. In fact, we have verified 
(data not presented here) that these features survive even in the presence of 
nearest-neighbour interactions among the hardcore particles. This leaves open the possibility of finding these features even in more realistic systems.

\section{Acknowledgements}

We thank S. S. Manna for very useful discussions. R.C. acknowledges the 
support from DAE (India) and DGAPA/UNAM (Mexico) postdoctoral fellowships.


\begin{thebibliography}{99}

\bibitem{Thouless} D. J. Thouless, Phys. Rev. B {\bf 27}, 6083 (1983).

\bibitem{Switkes} M. Switkes {\it et al.}, Science {\bf 283}, 1905 (1999); P. Leek {\it et al.}, Phys. Rev. Lett. {\bf 95}, 256802 (2005).

\bibitem{Watson} S. Watson {\it et al.}, Phys. Rev. Lett. {\bf 91}, 258301 (2003).

\bibitem{Astumian} R. D. Astumian, Phys. Rev. Lett. 91, 118102 (2003); R. D.
Astumian and P. Hanggi, Phys. Today 55, 33 (2002).

\bibitem{Rahav} S. Rahav, J. Horowitz, and C. Jarzynski, Phys. Rev. Lett. {\bf 101}, 140602 (2008).

\bibitem{Marathe} R. Marathe, A. M. Jayannavar, and A. Dhar, Phys. Rev. E {\bf 75}, 030103 (2007).

\bibitem{Julicher} F. Julicher, A. Ajdari, and J. Prost, Rev. Mod. Phys. {\bf 69}, 1269 (1997); P. Reimann, Phys. Rep. {\bf 361}, 57 (2002).

\bibitem{Marchesoni} F. Marchesoni, Phys. Rev. Lett. 77, 2364 (1996).

\bibitem{Seifert} U. Seifert, Phys. Rev. Lett. {\bf 106}, 020601 (2011).

\bibitem{Dhar_PRL2007} K. Jain, R. Marathe, A. Chaudhuri, and A. Dhar, Phys. Rev. Lett. {\bf 99}, 190601 (2007).

\bibitem{Dhar_JSTAT2008} R. Marathe, K. Jain, and A. Dhar, J. Stat. Mech.(2008) P11014.

\bibitem{Dhar_EPL2011} D. Chaudhuri and A. Dhar, EPL {\bf 94}, 30006 (2011).

\bibitem{Dhar_PRE2015} D. Chaudhuri, A. Raju, and A. Dhar, Phys. Rev. E {\bf 91}, 050103 (2015).

\bibitem{Libchaber} A. Simon and A. Libchaber, Phys. Rev. Lett. {\bf 68}, 3375 (1992); L. P. Faucheux, G. Stolovitzky, and A. Libchaber, Phys. Rev. E {\bf 51}, 5239 (1995).

\bibitem{Quake_RMP2005} T. M. Squires and S. R. Quake, Rev. Mod. Phys. {\bf 77}, 977 (2005).

%\bibitem{Service_Science1998} R. F. Service, Science {\bf 282}, 399 (1998).

\bibitem{Lipowsky_Science1999} H. Gau, S. Herminghaus, P. Lenz, and R. Lipowsky, Science {\bf 283}, 46(1999).

\bibitem{Marr_Science2002} A. Terray, J. Oakey, and D. W. M. Marr, Science {\bf 296}, 1841 (2002).

\bibitem{Penna_Tarazona_JChemPhys2003} F. Penna and P. Tarazona, J. Chem. Phys. {\bf 119}, 1766 (2003). 

\bibitem{Tarazona_Marconi_JchemPhys2008} P. Tarazona and U. M. B. Marconi, J. Chem. Phys. {\bf 128}, 164704 (2008).

\bibitem{Zon1} R. van Zon and E. G. D. Cohen, Phys. Rev. E {\bf 67}, 046102 (2003).

\bibitem{Zon2} R. van Zon and E. G. D. Cohen, Phys. Rev. Lett. {\bf 91}, 110601 (2003).

\bibitem{Pal} A. Pal and S. Sabhapandit, Phys. Rev. E {\bf 87}, 022138 (2013).

\bibitem{Wang} G. M. Wang, E. M. Sevick, E. Mittag, D. J. Searles, and D. J. Evans, Phys. Rev. Lett. {\bf 89}, 050601 (2002).

\bibitem{Blickle} V. Blickle, T. Speck, L. Helden, U. Seifert, and C. Bechinger, Phys. Rev. Lett. {\bf 96}, 070603 (2006). V. Blickle, T. Speck, C. Lutz, U. Seifert, and C. Bechinger, Phys. Rev. Lett. {\bf 98}, 210601 (2007).

\bibitem{Bechinger} T. Bohlein and C. Bechinger, Phys. Rev. Lett. {\bf 109}, 058301 (2012).

\bibitem{Ciliberto} J. R. Gomez-Solano, A. Petrosyan, S. Ciliberto, R. Chetrite, and K. Gawedzki, Phys. Rev. Lett. {\bf 103}, 040601 (2009). J. R. Gomez-Solano, L. Bellon, A. Petrosyan, and S. Ciliberto, EPL {\bf 89}, 60003 (2010).

\bibitem{Spitzer} F. Spitzer, Adv. Math. {\bf 5}, 246 (1970).

\bibitem{reviews} T. M. Liggett, {\it Interacting Particle Systems, Springer}, New York (1985). G. M. Schütz, in {\it Phase Transitions and Critical Phenomena}, edited by C. Domb and J. Lebowitz
(Academic, New York, 2000), pp. 3 - 242. 

\bibitem{Spohn} H. Spohn, J. Phys. A {\bf 16} 4275 (1983).

\bibitem{Schutz} G. Schutz and S. Sandow, Phys. Rev. E {\bf 49}, 2726 (1994).

%\bibitem{Brzank} A. Brzank, Diffusion Fundamentals {\bf 4}, 7.1 - 7.12 (2006).

\bibitem{Santos} J. E. Santos and G. M. Schutz, Phys. Rev. E {\bf 64}, 036107 (2001).

\bibitem{Derrida_2001} B. Derrida, J. L. Lebowitz, and E. R. Speer, Phys. Rev. Lett. {\bf 87}, 150601 (2001); T. Bodineau and B. Derrida, Phys. Rev. Lett. {\bf 92}, 180601 (2004).

%\bibitem{Bodineau} T. Bodineau, B. Derrida, J. L. Lebowitz, J. Stat. Phys. {\bf 131}, 821 (2008).

\bibitem{Karger} J. Karger and D.M. Ruthven, {\it Diffusion in Zeolites and Other Microporous Solids}, Wiley New York 1992. A. Brzank, Diffusion Fundamentals {\bf 4}, 7.1 - 7.12 (2006); A. Brzank and G.M. Sch\"utz, Diffus. Fund. 4, 7.1 (2006).

\bibitem{Nagar} A. Nagar, M. Ha, and H. Park, Phys. Rev. E {\bf 77}, 061118 (2008).

\bibitem{Rolland} C. Appert-Rolland, B. Derrida, V. Lecomte, and F. van Wijland, Phys. Rev. E {\bf 78}, 021122 (2008).

%\bibitem{Derrida2001} B. Derrida, J. L. Lebowitz, and E. R. Speer, Phys. Rev. Lett. {\bf 87}, 150601 (2001).

%\bibitem{Bodineau2010} T. Bodineau, B. Derrida, J. L. Lebowitz, J. Stat. Phys. {\bf 140}, 648 (2010).

\bibitem{Sadhu} T. Sadhu, S. N. Majumdar, and D. Mukamel, Phys. Rev. E {\bf 84}, 051136 (2011).

\bibitem{Hegde} C. Hegde, S. Sabhapandit, and A. Dhar, Phys. Rev. Lett. {\bf 113}, 120601 (2014).

\bibitem{Mallick} C. Arita, P. L. Krapivsky, and K. Mallick, Phys. Rev. E {\bf 90}, 052108 (2014).

\bibitem{Chatterjee2014} R. Chatterjee, S. Chatterjee, P. Pradhan, and S. S. Manna, Phys. Rev. E {\bf 89}, 022138 (2014).

\bibitem{Rajewsky} N. Rajewsky, L. Santen, A. Schadschneider, and M. Schreckenberg, J. Stat. Phys. {\bf 92}, 151 (1998).


\bibitem{kol2} A. B. Kolomeisky, G. M. Sch ̈utz, E. B. Kolomeisky, and J. P.
Straley, J. Phys. A: Math. Gen. 31, 6911 (1998)
\bibitem{dudz2} M. Dudzinsky and G. M. Sch ̈utz, J. Phys. A: Math. Gen. 33,
8351 (2000).
\bibitem{santen2} L. Santen and C. Appert, J. Stat. Phys. 106, 187 (2002).
\bibitem{popkov2} V. Popkov and G. M. Sch ̈utz, Europhys. Lett. 48, 257 (1999).







\end{thebibliography}
\end{document}